\documentclass[11pt,a4paper]{article}
\usepackage[T1]{fontenc}
\usepackage{mathpazo}
\usepackage[margin=27mm]{geometry}
\usepackage{amsmath,amssymb,amsthm,mathtools}
\usepackage{booktabs,array,enumitem,microtype}
\usepackage{tikz}
\usetikzlibrary{arrows.meta,positioning,decorations.pathreplacing}
\usepackage[hidelinks]{hyperref}
\hypersetup{pdftitle={Concave Run Budgets and Restart Width},pdfauthor={Nassim Dehouche}}
\setlist{itemsep=2pt,topsep=4pt}
\newtheorem{theorem}{Theorem}[section]
\newtheorem{lemma}[theorem]{Lemma}
\newtheorem{proposition}[theorem]{Proposition}
\newtheorem{corollary}[theorem]{Corollary}
\theoremstyle{definition}
\newtheorem{definition}[theorem]{Definition}
\newtheorem{example}[theorem]{Example}
\theoremstyle{remark}
\newtheorem{remark}[theorem]{Remark}

\newcommand{\N}{\mathbb{N}}
\newcommand{\Z}{\mathbb{Z}}
\newcommand{\Q}{\mathbb{Q}}
\newcommand{\poly}{\operatorname{poly}}
\newcommand{\lab}{\operatorname{lab}}
\newcommand{\last}{\operatorname{last}}
\newcommand{\wl}{w}
\renewcommand{\wr}{w_{\mathrm r}}
\newcommand{\Lr}{\Lambda_{\mathrm r}}
\newcommand{\W}[1]{\mathrm{W}[#1]}
\newcommand{\FPT}{\mathrm{FPT}}
\newcommand{\XP}{\mathrm{XP}}
\newcommand{\NP}{\mathrm{NP}}
\newcommand{\OPT}{\mathrm{OPT}}
\newcommand{\qrep}{q_{\mathrm{rep}}}
\newcommand{\blk}{\mathsf{N}}
\newcommand{\eps}{\varepsilon}

\title{\textbf{Concave Run Budgets and Restart Width}}
\author{Nassim Dehouche\thanks{Mahidol University International College, Mahidol University, 73170, Salaya Thailand}}
\date{November 2026}

\begin{document}
\maketitle

\begin{abstract}
A source-to-target path in an edge-labelled DAG uses each label in some number of maximal runs. A label used in $r$ runs is charged $f(r)$, where $f$ is a nondecreasing concave rational tariff, and we ask for a path of total charge at most~$B$, or for a shortest such path. We parameterize by \emph{restart width} $\wr$: the largest number of labels that, at some vertex, have already been used and can still start a new run. Restart width never exceeds the number of labels occurring both before and after a vertex (\emph{live width}), and can be arbitrarily smaller.

We establish tractability and hardness criteria in terms of the number $J_f(T)$ of distinct marginal rates of $f$ up to horizon~$T$. For minimum charge and feasibility, a dynamic program that commits lazily to affine charging plans runs in time $J_f(T)^{\wr}\poly(N)$; it handles arbitrary per-label tariffs and needs no bound on the numerical size of the charges, so the problem is fixed-parameter tractable whenever $J_f$ is effectively subpolynomial. Conversely, for every fixed tariff with $J_f(T)\ge T^{\delta}$ eventually, feasibility is $\NP$-complete and $\W1$-hard in~$\wr$, already on planar bipartite DAGs of maximum degree three and treewidth two in which all paths have the same length, and it admits no $g(\wr)N^{o(\wr)}$ algorithm under ETH. The reduction enforces consistency through pairwise Jensen gaps that are realised by positive charges only; it therefore needs no assumption on charge denominators and covers, for instance, harmonic, geometric and power-law tariffs uniformly. A $k$-colour clique instance has width at most $2k+1$, and a residue-compensation variant lowers this to $k+1$ when denominators are polynomially bounded. As a consequence there are tariffs $F\le F_M\le(1+2/M)F$ for which the problem is $\W1$-hard for $F$ and fixed-parameter tractable for every~$F_M$. We also give an FPT budget-relaxing scheme and an accuracy obstruction, hardness in the total number of labels under disjunctive truth-table reductions on polynomial grids, $\W{\mathrm P}$-membership for that parameter, and a separate numerical source of hardness for binary-encoded lengths. Oscillating plan growth and non-effective subpolynomial bounds remain unresolved.
\end{abstract}

\section{Introduction}\label{sec:intro}

Many resources are paid for per \emph{episode of use} rather than per unit of use, and repeated episodes are discounted. A transit pass prices the second boarding of a line lower than the first and caps the total; a machine has a set-up cost each time it is switched back on, and maintenance contracts make later set-ups cheaper; a licence is reactivated at a declining fee. The common structure is a sequence of activities, each carried out with one resource, in which staying with a resource is free, returning to it starts a new episode, and the charge of a resource is a concave function of its number of episodes.

We study the routing problem this structure defines. The input is a directed acyclic graph with a source and a target whose edges carry labels (the resources). On a source-to-target path~$P$, a \emph{run} of a label~$c$ is a maximal block of consecutive $c$-labelled edges, and $r_c(P)$ denotes the number of runs of~$c$. Given a nondecreasing concave \emph{tariff} $f$ with $f(0)=0$, the path is charged
\[
  \Phi(P)=\sum_{c} f\bigl(r_c(P)\bigr).
\]
\textsc{Concave Run Budget} asks whether some path has $\Phi(P)\le B$; its optimisation variant asks for a shortest such path with respect to independent nonnegative edge lengths. For $f(r)=r$ the charge is the number of label changes plus one, and for $f(r)=\min\{r,1\}$ it is the number of distinct labels used, so the problem interpolates between two classical objectives. Minimising the number of line changes is polynomial on a last-label product graph~\cite{BohmovaEtAl18,Dehouche20}, whereas minimising the number of distinct labels is the minimum-label path problem, which is $\NP$-hard, hard to approximate~\cite{BroersmaLWZ05,HassinMS07,BohmovaEtAl18} and $\W2$-hard in the number of labels used~\cite{FellowsGK10}. The tariff expression is already encompassed by the crossing-cost frameworks of Blanco et al.~\cite{BlancoEtAl16,BlancoEtAl17}, as explained in Section~\ref{sec:related}. Our contribution is a parameterized analysis of this family in terms of restart width and marginal-rate growth, with a delayed-commitment algorithm and positive-charge consistency reductions.

What makes the general problem hard is memory: the charge of the next run of~$c$ depends on how many runs of~$c$ the path has already made. Only some of this history matters at a given vertex. A label that cannot occur again may be forgotten, and so may a label that can occur again but only as a continuation of its current run. We call the number of labels whose history genuinely has to be remembered at a vertex its \emph{restart width} and write $\wr$ for the maximum over all vertices (Definition~\ref{def:restart}). Restart width is bounded by the \emph{live width} $\wl$, the maximum number of labels that occur both on some prefix and on some continuation at a vertex, and both are small when resources are localised in time; Section~\ref{sec:sources} lists simple structural conditions. The obvious dynamic program stores a run count for every remembered label and takes time $(T+1)^{\wr}\poly(N)$, where $T$ is the largest number of edges on a path. The question of this paper is for which tariffs the exponent can be removed.

\subsection{Results}

Let $\Delta_j=f(j)-f(j-1)$ be the marginal rates of~$f$ and let $J_f(T)$ be the number of distinct values among $\Delta_1,\dots,\Delta_T$. We call $f$ \emph{effectively subpolynomial} if there is a computable $\tau$ with $J_f(T)\le T^{1/p}$ for all integers $p\ge1$ and $T\ge\tau(p)$, and we write
\[
  L_f=\liminf_{T\to\infty}\frac{\log J_f(T)}{\log T},\qquad
  U_f=\limsup_{T\to\infty}\frac{\log J_f(T)}{\log T},
\]
so that $L_f>0$ means $J_f(T)\ge T^{\delta}$ for some $\delta>0$ and all sufficiently large~$T$. Throughout, $N$ is the input size and a fixed tariff is assumed to be rational and exactly computable up to horizon $T$ in time polynomial in~$T$.

\begin{theorem}[Main theorem]\label{thm:main}
Let $f$ be a fixed nondecreasing concave rational tariff with $f(0)=0$.
\begin{enumerate}[label=\textup{(\arabic*)}]
\item \textsc{Concave Run Budget} can be solved in time $J_f(T)^{\wr}\poly(N)$. Hence it is fixed-parameter tractable in $\wr$ if $J_f$ is effectively subpolynomial. The same holds for shortest feasible paths with polynomially bounded integer lengths, and for per-label tariffs with $J=\max_c J_{f_c}$.
\item If $L_f>0$, then \textsc{Concave Run Budget} is $\NP$-complete and $\W1$-hard parameterized by $\wr$, and also by $\wl$, even on simple planar bipartite DAGs of maximum degree three and treewidth at most two in which all source-to-target paths have the same number of unit-length edges. A $k$-colour instance of \textsc{Multicolored Clique} is mapped to an instance with $\wr\le\wl\le 2k+1$. Unless ETH fails, there is no algorithm with running time $g(\wr)N^{o(\wr)}$ for any computable~$g$.
\item If moreover all values $f(0),\dots,f(T)$ lie in $\frac{1}{D_f(T)}\Z$ for a polynomially bounded $D_f$, the reduction in \textup{(2)} can be made to have $\wr\le\wl\le k+1$.
\end{enumerate}
\end{theorem}

Part~(1) is Theorem~\ref{thm:plans}, part~(2) is Theorem~\ref{thm:hard} and part~(3) is Theorem~\ref{thm:residue}. Neither (1) nor (2) depends on the numerical size of the charges: the algorithm compares exact rationals and the reduction uses tariff values only as they come. The two conditions are complementary up to oscillation. Since $1\le J_f(T)\le T$, a tariff has $U_f=0$ exactly when $J_f(T)=T^{o(1)}$, and the only tariffs not classified are those with $L_f=0<U_f$, whose plan count is polynomial along some horizons and subpolynomial along others, together with non-effective instances of $U_f=0$ (Section~\ref{sec:open}).

The growth of the \emph{value} of the tariff is irrelevant; what matters is the number of distinct rates. Section~\ref{sec:examples} exhibits a strictly concave tariff $F$ of logarithmic growth and, for every integer $M\ge1$, a tariff $F_M$ with
\[
   F\le F_M\le\Bigl(1+\frac2M\Bigr)F,
\]
such that the problem is $\W1$-hard for~$F$ and fixed-parameter tractable for every $F_M$ (Proposition~\ref{prop:twins}). Tariffs on a fixed monetary grid, finite menus of plans and caps are always on the tractable side (Proposition~\ref{prop:grid}); hardness needs marginal rates of ever finer resolution.

We complement the main theorem in three directions.
\begin{itemize}
\item \emph{Budget relaxation.} For arbitrary per-label tariffs and arbitrary rational lengths, a path no longer than the optimum and of charge at most $(1+\eps)B$ can be found in time $g(\wr,1/\eps)\poly(N)$ (Theorem~\ref{thm:relax}). For hard tariffs with inverse-polynomial separation the dependence on $1/\eps$ cannot be made polynomial unless $\FPT=\W1$ (Corollary~\ref{cor:accuracy}).
\item \emph{Few labels.} The total number $q$ of labels is a weaker parameter than width. For tariffs with $L_f>0$ on a polynomial grid the problem remains $\W1$-hard in $q$ under disjunctive truth-table reductions, with $2k+1$ labels, and for arbitrary per-label tariffs it belongs to $\W{\mathrm P}$ (Section~\ref{sec:few}).
\item \emph{Binary lengths.} With a length bound in binary, the harmonic tariff is $\NP$-hard already at live width one, by a number-theoretic encoding unrelated to width (Proposition~\ref{prop:harmonic}); tariffs on a polynomial grid do not show this effect.
\end{itemize}

\subsection{Techniques}

\paragraph{Algorithm.} A concave tariff is the lower envelope of the affine functions $\beta_j+\Delta_j r$ that support it; each is a \emph{charging plan} with a set-up fee and a per-run rate. This representation of concave costs is standard~\cite{MagnantiS04,MagnantiS12}. A path may guess, for each label, the plan that will be tight for its final run count, and pay according to that plan. We observe that the guess can be postponed: a label pays the default rate $f(1)$ until it commits, and committing late never undercuts the tariff (Lemma~\ref{lem:delayed}). The state of a label is then its committed plan or ``default'', and unused labels and uncommitted labels are indistinguishable. Together with forgetting outside the restart set this gives $J_f(T)^{\wr}$ states per vertex and last label.

\paragraph{Hardness.} We reduce from \textsc{Multicolored Clique}. A selection gadget picks a vertex in each colour class and an edge for each pair of classes; the budget must force every picked edge to end at the picked vertices. Encode the index $x$ of a vertex by a run count $r_x$. If two gadgets give a shared label $r_x$ and $r_y$ runs respectively, its charge is $f(r_x+r_y)$, and concavity gives
\[
   2f(r_x+r_y)-f(2r_x)-f(2r_y)\;\ge\;0 .
\]
Equality holds if $x=y$, and for $x\ne y$ it fails as soon as $f$ has a strict change of rate between $2r_x$ and $2r_y$. So we use \emph{two} shared labels for the pair of gadgets, and let each gadget pay, on private labels, a constant minus $f(2r_x)$. The latter needs no subtraction: with $z$ candidate indices, the gadget selecting $x$ runs through private labels with $2r_y$ runs for every $y\ne x$, which costs $\sum_{y}f(2r_y)-f(2r_x)$ up to a constant. All charges are tariff values, so no arithmetic assumption on $f$ is needed. Codes $r_1<\dots<r_z$ of polynomial size with a rate change between consecutive doubled values exist exactly when $L_f>0$ (Lemma~\ref{lem:codes}). Checking each colour along a chain of its $k$ gadgets keeps two shared labels per colour alive at any time, whence width $2k+1$.

With a single shared label per colour and $k$ gadgets one would use the $k$-point Jensen inequality instead and must realise the charge $C-f(kr_x)/k$, which is not a combination of tariff values. On polynomial grids this can still be done with positive charges after restricting the code to one residue class modulo~$k$ (Section~\ref{sec:residue}); this is how part~(3) of the main theorem saves a factor of two in the width.

\subsection{Related work}\label{sec:related}

\emph{Labels, colours and changes on paths.} The two extreme tariffs are classical. Paths with few distinct labels are studied in~\cite{BroersmaLWZ05,HassinMS07}, with parameterized results for the number of used labels and for graph width measures in~\cite{FellowsGK10}. B\"ohmov\'a et al.~\cite{BohmovaEtAl18} consider routes in transit networks whose colours are lines: a route with fewest line changes is found in polynomial time, while minimising the number of distinct lines is hard to approximate; the same two constraints define the interchange- and colour-constrained shortest paths of~\cite{Dehouche20}, and the number of boardings is the standard second criterion of transit routing~\cite{DellingPW15}. Davot, Giroudeau and K\"onig~\cite{DavotGK24} study both objectives when every arc carries a list of colours and obtain $\W2$-hardness and an algorithm that is fixed-parameter in the total number of colours. Bentert, Kellerhals and Niedermeier~\cite{BentertKN26} bound the number of \emph{occurrences} of each colour on a short path and prove $\W1$-hardness in the number of colours by a reduction from \textsc{Multicolored Clique} with parallel-path selection gadgets, a template that our reductions share. Their separate study of balance-fair shortest paths~\cite{BentertKN22} classifies several structural parameterizations. These colour-count constraints differ from a single budget on a sum of concave run tariffs.

\emph{Reload and changeover costs.} These costs depend on the two colours involved in a transition~\cite{WirthS01,GamvrosGR12,GourvesLMM10}. A last-colour product graph handles nonnegative transition costs for walks, and for paths in a DAG. It does not in general enforce simplicity in a cyclic graph or solve spanning-tree and arborescence variants. The latter include minimum changeover cost arborescence, with distinct tractability and hardness results in treewidth and degree~\cite{GozupekSSZ16,GozupekOPSS17}. Our additional source of history is that entering a label costs an amount depending on its earlier run count.

\emph{Crossing costs.} The general model of Blanco et al.~\cite[Section~2.1]{BlancoEtAl17} contains our tariff expression. Extend each tariff linearly between integers, take each label's edges as a region, and use the discrete metric. Every nonempty DAG run has distinct endpoints, so its regional charge is $f_c(r_c(P))$. Their arc weights are $\varphi d(u,v)$ for a positive constant~$\varphi$; hence their full objective is
\[
                 C_{\varphi}(P)=\varphi |P|+\Phi(P).
\]
Consequently, minimum charge is recovered on equal-length instances. More generally, if all tariff values up to $T$ lie in $D^{-1}\Z$, choose $0<\varphi<1/(D(T+1))$: the additive perturbation is smaller than any positive charge gap, so every minimizer of $C_{\varphi}$ minimizes $\Phi$. This is a deduction from their model. Our \textsc{Shortest Budgeted Path}, with independent lengths and a hard charge budget, is not a direct special case of that scalar objective.

The earlier ZIB report~\cite[pp.~8, 11--15, 17--20]{BlancoEtAl16} studies linear, constant and piecewise-constant regional functions. Its Theorem~1 treats disjoint linear regions, Theorem~3 treats laminar linear regions, and Proposition~2 allows at most two linear regions per arc; Proposition~3 concerns flat fees, and Theorem~2 gives an FPTAS for one nontrivial piecewise-constant region measured by additive distance. Appendix~A.2 retains entry vertices of currently traversed regions. Our label regions are disjoint, but nonlinear run charges require remembering a label after leaving it; simultaneous region overlap does not bound that history. These results do not give the restart-width and marginal-growth criteria studied here. The 2017 polynomial case assumes linear functions and pairwise disjoint regions~\cite[Section~2.2]{BlancoEtAl17}.

\emph{Fares and resource-constrained shortest paths.} Our optimisation variant is a shortest path problem with one non-additive resource. Labelling algorithms for such problems are surveyed in~\cite{IrnichD05}, and Parmentier~\cite{Parmentier19} treats non-linear resources in a lattice-ordered monoid. Fare structures of public transport are modelled by monoids of tickets and conditional fare networks by Euler, Lindner and Bornd\"orfer~\cite{EulerB19,EulerLB24}, and Sch\"obel and Urban~\cite{SchobelU22} classify the cheapest ticket problem for distance- and zone-based fares into polynomial and $\NP$-complete cases. These works address dominance, modelling power and classical complexity. We ask when one specific history, the run counts, can be compressed to a number of states that is fixed-parameter bounded. The numerical side of the two-criteria problem is that of restricted shortest paths~\cite{Hassin92,LorenzR01,TsaggourisZ09}, from which the budget-relaxing scheme takes its rounding.

\emph{Concave costs and plans.} Goel, Tripathi and Wang~\cite{GoelTW10} study combinatorial problems with concave discounted prices, including shortest paths in undirected graphs: selected edges are allocated to agents, each applying a concave function to its total assigned cost. Our labels are fixed and their charges depend on maximal runs. Zhang, Shen and Song~\cite{ZhangSS16} study a bi-attribute shortest-path objective with a concave function of an aggregate path attribute. Count-dependent concave penalties also occur in degree-constrained orientation~\cite{AsahiroJMO22} and set cover~\cite{CardinalD08}; formal-language constraints~\cite{BarrettJM00} and nonadditive traffic-assignment costs~\cite{GabrielB97} are further relatives.

Writing a concave cost as a minimum of fixed-charge affine pieces is standard~\cite{MagnantiS04,MagnantiS12}, as is approximating it by a small piecewise-linear menu~\cite{MagnantiS04}. Our algorithm uses this representation with delayed commitment, merges unused and default states, and forgets plans outside the restart set. Bredereck et al.~\cite[Section~2, Theorem~2]{BredereckEtAl20} obtain fixed-parameter algorithms for mixed integer programs with few integer variables and constraints of the form $\sum_i F_i(x_i)\le\sum_i G_i(x_i)+b$, with convex $F_i$ and concave $G_i$. Each such inequality defines a convex continuous feasible set. An upper bound on a sum of concave costs has the opposite orientation and is generally nonconvex, so their theorem does not directly apply to our budget constraint. Theorem~\ref{thm:main} instead relates the growth of the plan count to tractability in restart width.

\emph{Width parameters.} Live width is a cut measure on label occurrences. On a single path, associate with each label the interval of internal cuts after its first edge and before its last, omitting empty intervals. Then live width is the clique number of the auxiliary interval graph on these label lifetimes and, when this graph is nonempty, one more than its pathwidth. This is an observation about the auxiliary graph, not the underlying path; Kinnersley~\cite{Kinnersley92} establishes the general equality between vertex separation and pathwidth. The vertex-interval-membership width of temporal graphs, introduced by Bumpus and Meeks~\cite{BumpusM23} and developed further in~\cite{EnrightHLM26}, uses the related idea of vertices active between their first and last time-edge. In a branching DAG, liveness is taken over all prefixes and continuations of a vertex, as in the live-variable analysis of compilers, and restart width refines it further. Our contribution is the use of these label-history sets in the plan algorithm and hardness classification. Live width is incomparable with the treewidth of the underlying graph: the path with label word $c_1\cdots c_qc_1\cdots c_q$ has live width $q$, and an acyclic tournament with pairwise distinct edge labels has live width zero. Our hardness results hold on graphs of treewidth two.

\subsection{Organisation}
Section~\ref{sec:model} fixes the model, defines the two width parameters and proves the forgetting lemma. Section~\ref{sec:alg} gives the plan algorithm, Section~\ref{sec:hard} the lower bound and Section~\ref{sec:residue} its refinement on polynomial grids. Section~\ref{sec:examples} discusses concrete tariffs and separation gaps. Sections~\ref{sec:relax}, \ref{sec:few} and~\ref{sec:binary} treat budget relaxation, few labels and binary lengths, and Section~\ref{sec:open} lists open problems.

\section{Model and width parameters}\label{sec:model}

\subsection{Tariffs and problems}

A \emph{tariff} is a function $f\colon\N\to\Q_{\ge0}$ with $f(0)=0$ whose \emph{marginals} $\Delta_j=f(j)-f(j-1)$, $j\ge1$, satisfy $\Delta_1\ge\Delta_2\ge\dots\ge0$; that is, $f$ is nondecreasing and concave. An integer $t\ge1$ is a \emph{drop} of $f$ if $\Delta_t>\Delta_{t+1}$, and $d_t=\Delta_t-\Delta_{t+1}\ge0$ denotes the size of the decrement at~$t$. For a horizon $T\ge1$,
\[
  J_f(T)=\bigl|\{\Delta_1,\dots,\Delta_T\}\bigr| = 1+\bigl|\{t<T: t\text{ is a drop}\}\bigr| ,
\]
and $D_f(T)$ denotes the least positive integer $D$ with $f(0),\dots,f(T)\in\frac1D\Z$. We say that $f$ lies on a \emph{polynomial grid} if $D_f(T)\le T^{c}$ for a constant $c$ and all $T\ge 2$.

An instance consists of a DAG $G=(V,E)$ with a source $s$ and a target $t$, a label $c(e)$ for every edge, and a budget $B\in\Q_{\ge0}$. We assume that $G$ is \emph{trimmed}: every vertex and edge lies on an $s$--$t$ path. A \emph{path} is an $s$--$t$ path unless said otherwise, a \emph{prefix at $v$} is an $s$--$v$ path and a \emph{continuation at $v$} is a $v$--$t$ path. For a walk $P$ we write $\lab(P)$ for its set of labels, $\last(P)$ for the label of its last edge, and $r_c(P)$ for its number of maximal runs of~$c$. We let $T\ge1$ be the largest number of edges of a path and $q$ the number of labels.

In the \emph{common-tariff} model one tariff $f$ is fixed in advance and is not part of the input. We assume that $f(0),\dots,f(T)$ can be computed in time polynomial in $T$; in particular these numbers have polynomially many bits, although $D_f(T)$ may be exponential. In the \emph{per-label} model every label $c$ has its own tariff $f_c$, given in the input by a representation from which $f_c(0),\dots,f_c(T)$ can be computed in time polynomial in the input size, such as a table of marginals, a finite menu of affine plans, or a cap $f_c(r)=\min\{a_cr,b_c\}$ with binary coefficients. The charge of a path is $\Phi(P)=\sum_c f_c(r_c(P))$. All algorithms are stated for the per-label model and all lower bounds for the common-tariff model. In both, $N$ denotes the total bit length of the input.

\begin{description}[leftmargin=1.5em,itemsep=1pt]
\item[\textsc{Concave Run Budget}:] decide whether some path $P$ has $\Phi(P)\le B$.
\item[\textsc{Shortest Budgeted Path}:] given in addition nonnegative rational edge lengths $\ell$, find a path minimising $\ell(P)$ subject to $\Phi(P)\le B$.
\end{description}
We refer to the first problem as \emph{feasibility}. Hardness results always concern feasibility. For $\ell$ we distinguish unit lengths, integer lengths bounded by a given $W$, and arbitrary rational lengths in binary.

By a fixed-parameter algorithm we mean one with running time $g(p)\poly(N)$ for a \emph{computable} $g$; this is the reason for the word ``effectively'' in the definition of effectively subpolynomial tariffs. If only $J_f(T)=T^{o(1)}$ is known, the algorithm of Section~\ref{sec:alg} still runs in time $g(\wr)\poly(N)$ for some function~$g$, which need not be computable; in the terminology of Downey and Fellows~\cite[Chapter~2, Definitions~2.1.1 and~2.2.1(i)]{DowneyF13} the problem is then uniformly but perhaps not strongly uniformly fixed-parameter tractable.

\begin{remark}[Cyclic networks and policies]\label{rem:walks}
Deleting symbols from a word does not increase any run count: map each $c$-run of the remaining word to the $c$-run of the original word that contains its first symbol; two runs with the same image would be separated by a remaining symbol different from $c$ inside one run of $c$. Hence, in a cyclic network, possibly in product with a finite automaton restricting the admissible label sequences~\cite{BarrettJM00}, closed subwalks can be removed without increasing charge or length, and it suffices to consider walks with fewer edges than the product has vertices. Unrolling the product to that horizon gives a DAG to which our algorithms apply. Unrolling may increase the width considerably, since every label of a strongly connected component can be live in the middle layers; the parameters are meaningful primarily for acyclic or time-expanded inputs.
\end{remark}

\subsection{Live width and restart width}

For a vertex $v$ let $\Pi(v)$ be the set of labels occurring on some prefix at $v$ and $\Gamma(v)$ the set of labels occurring on some continuation at~$v$.

The restart set can be derived directly from the run-count update rule. Keep the last label to decide whether the next edge starts a run. A label's older history can affect a future payment only if it has already occurred and some continuation can start another run of it. Taking this condition over all prefixes and continuations gives the following graph-defined set; the projection construction below proves that it is sufficient for exact dynamic programming.

\begin{definition}\label{def:restart}
The \emph{live set} of $v$ is $\Lambda(v)=\Pi(v)\cap\Gamma(v)$, and the \emph{live width} of the instance is $\wl=\max_v|\Lambda(v)|$. The \emph{restart set} of $v$ is
\[
  \Lr(v)=\bigl\{c:\ c\in\lab(P)\text{ and } r_c(PQ)>r_c(P)\text{ for some prefix $P$ and continuation $Q$ at $v$}\bigr\},
\]
and the \emph{restart width} is $\wr=\max_v|\Lr(v)|$.
\end{definition}

Both definitions quantify over all paths of the graph, irrespective of the budget. Clearly $\Lr(v)\subseteq\Lambda(v)$ and $\wr\le\wl$. A label that occurs on a single edge is never live, because a prefix and a continuation at $v$ concatenate to a path. Hence $\wl\le\qrep\le q$, where $\qrep$ is the number of labels occurring on at least two edges.

\begin{lemma}\label{lem:restart}
\begin{enumerate}[label=\textup{(\roman*)}]
\item Let $P$ be a prefix at $v$ and $Q$ a continuation at $v$. Whether an edge of $Q$ starts a new run in $PQ$ depends only on $Q$ and $\last(P)$.
\item The sets $\Lr(v)$ can be computed in polynomial time.
\item For every $n\ge1$ there is a simple DAG with $\wl=n$ and $\wr=0$.
\end{enumerate}
\end{lemma}
\begin{proof}
(i) An edge starts a new run exactly if the edge preceding it on $PQ$ carries a different label or does not exist.

(ii) For a label $c$ let $A_c(v)$ hold if some prefix at $v$ contains $c$ and ends with a label other than~$c$, and let $Z_c(v)$ hold if some continuation at $v$ contains a $c$-edge immediately preceded, within the continuation, by an edge with another label. By (i), a run of $c$ is started in $Q$ after $P$ either at the first edge of~$Q$, which requires $\last(P)\ne c$, or at a later edge, and conversely if $\last(P)\neq c$ and $c\in\lab(Q)$ then the first $c$-edge of $Q$ starts a run. Therefore
\[
   c\in\Lr(v)\iff \bigl(A_c(v)\wedge c\in\Gamma(v)\bigr)\ \vee\ \bigl(c\in\Pi(v)\wedge Z_c(v)\bigr).
\]
Each predicate is a reachability question in the product of $G$ with a three-state automaton and is computed by one pass in topological order.

(iii) Take $n$ internally disjoint two-edge paths from $s$ to a vertex $v$ and $n$ internally disjoint two-edge paths from $v$ to $t$; label both edges of the $i$-th path on either side with~$c_i$. Then $\Lambda(v)=\{c_1,\dots,c_n\}$. Every path has the word $c_ic_ic_jc_j$, in which no label starts a second run, so all restart sets are empty.
\end{proof}

We now define the projection explicitly. For each label $c$, let $H_c$ be a state set with initial state $\bot_c$, let $A_c(h)$ be the admissible choices in state $h$, and let $U_c(h,a)$ and $K_c(h,a)$ be the updated state and payment. The full process starts with $h(c)=\bot_c$ for every $c$. At a run start of $c$ it chooses $a\in A_c(h(c))$, pays $K_c(h(c),a)$ and replaces $h(c)$ by $U_c(h(c),a)$. On a continuation edge it pays zero and changes no history state. Other labels are unchanged at every step.

The \emph{projected key} at $v$ is $(v,b,\sigma)$, where $b$ is the last label (or an initial symbol $\star$) and $\sigma(c)\in H_c$ is stored for $c\in\Lr(v)$. Given an edge $e=(v,u)$ of label $c$, set
\[
  \bar\sigma(d)=\begin{cases}
    \sigma(d),&d\in\Lr(v),\\
    \bot_d,&d\notin\Lr(v).
  \end{cases}
\]
If $b\ne c$, choose $a\in A_c(\bar\sigma(c))$, pay $K_c(\bar\sigma(c),a)$ and let $\widehat\sigma$ equal $\bar\sigma$ except that $\widehat\sigma(c)=U_c(\bar\sigma(c),a)$. If $b=c$, pay zero and set $\widehat\sigma=\bar\sigma$. The next key is
\[
                    (u,c,\widehat\sigma|_{\Lr(u)}).
\]
Initially all stored states are $\bot_c$. Thus admissible choices, payment and next key are functions only of the current key, edge and choice; no unrecorded prefix is consulted.

\begin{lemma}[Projection]\label{lem:projection}
For every source-to-target path, the full and projected processes have the same admissible choice sequences and the same payments along each such sequence.
\end{lemma}
\begin{proof}
Couple the processes along a path. For every prefix $P$ ending at $v$ and every $d\in\Lr(v)$ maintain
\begin{equation}\label{eq:inv}
 \sigma(d)=h(d)\quad\text{or}\quad
 \bigl(d\in\lab(P)\text{ and no continuation at $v$ starts a new run of $d$ after $P$}\bigr).
\end{equation}
The invariant holds initially. Suppose $e=(v,u)$ has label $c$ and starts a run. If $c\in\Lr(v)$, the second alternative is impossible, since a continuation through $e$ starts a run of $c$; hence $\bar\sigma(c)=h(c)$. If $c\notin\Lr(v)$, then $c\notin\lab(P)$, for otherwise $P$ and a continuation through $e$ would witness membership in $\Lr(v)$. Again $\bar\sigma(c)=h(c)=\bot_c$. The choices, payment and update therefore agree. On a continuation edge both processes do nothing.

To propagate the invariant, fix $d\in\Lr(u)$. If $e$ starts a run of $d$, both updated states agree by the preceding argument. Otherwise neither process updates $d$. If $d\in\Lr(v)$, equal old states remain equal; the second alternative, when applicable, persists because every continuation $Q$ at $u$ gives the continuation $eQ$ at $v$. If $d\notin\Lr(v)$ and $d\notin\lab(P)$, both old states are $\bot_d$. If $d\notin\Lr(v)$ but $d\in\lab(P)$, no continuation at $v$ can restart $d$ after $P$; applying this to $eQ$ proves the second alternative at $u$. This completes the induction. At every run start the state consulted by either process is the same, proving admissibility and equality of payments in both directions.
\end{proof}

A dynamic program on these keys is therefore exact. Lemma~\ref{lem:restart}(i) determines run starts from the last label, and the explicit transition rule determines the next key and payment. The last label is stored even when it is outside the restart set.

\begin{proposition}[Count-state baseline]\label{prop:baseline}
For per-label tariffs, a path of minimum charge can be computed in time $(T+1)^{\wr}\poly(N)$. If $\wr=0$, then every used label has exactly one run, $\Phi(P)=\sum_{c\in\lab(P)}f_c(1)$ is additive over run starts, and minimum-charge computation and feasibility are polynomial.
\end{proposition}
\begin{proof}
Apply Lemma~\ref{lem:projection} with $h(c)=r_c$ and cost $f_c(h(c)+1)-f_c(h(c))$ at a run start, and keep the minimum accumulated charge for each key. There are at most $|V|(q+1)(T+1)^{\wr}$ keys. If two runs of $c$ occurred on a path, $c$ would be in the restart set of the tail of the edge starting the second run.
\end{proof}

\subsection{Structural sources of small width}\label{sec:sources}

Since $\wr\le\wl$, any bound on live width is a bound on the parameter. Three simple situations give such bounds.

\emph{Temporal localisation.} Suppose vertices carry times that do not decrease along edges, and all endpoints of $c$-labelled edges have times in an interval $I_c$. A label that is live at $v$ has an edge before and an edge after $v$, so $I_c$ contains the time of~$v$. If no point of time lies in more than $\omega$ intervals, then $\wl\le\omega$. Validity windows of contracts in a time-expanded network are of this form.

\emph{Bounded recurrence.} In a layered DAG in which the edges leaving each layer use at most $b$ labels and the first and last layers in which a label is used are at most $h$ apart, a label live in layer $i$ is used in one of the $h$ preceding layers, so $\wl\le bh$.

\emph{Sessions.} If the DAG is a series composition of sessions, at most $a$ labels are used in more than one session and each session has at most $b$ further labels occurring on more than one edge, then $\wl\le a+b$, regardless of the number and size of the sessions.

\emph{Trips that cannot be re-boarded.} This example is intended in the per-label tariff model, so trips and contracts may have different tariffs. Consider a time-expanded timetable with the following single-label assignment. Movement and dwell arcs aboard trip $\tau$ carry its trip label $c_\tau$, so staying aboard through a stop continues the same run. Other arcs representing separately billed access or transfer services carry one of at most $a$ persistent contract labels, distinct from all trip labels; any remaining arcs have fresh labels. Thus each arc carries exactly one label, and a contract label never interrupts an on-vehicle trip segment. Suppose that, after leaving a trip, no feasible continuation can catch that trip again. Then trip labels never belong to a restart set, although they can be live along their routes. Only the $a$ contract labels can restart, giving $\wr\le a$ regardless of the number of live trip labels. This hypothesis is substantive: walking shortcuts or faster connecting services can make re-boarding possible.

A stylised pulse-transfer hub illustrates the separation in Lemma~\ref{lem:restart}(iii), with longer route segments and a shared source and target. Let $n$ through trips meet at one transfer event $v$, with one incoming and one outgoing route segment for each trip and no further interchanges. Collapse the transfer event to one vertex, so staying aboard keeps the same label. Every route has label word $c_i^{a_i}c_j^{b_j}$ for positive $a_i,b_j$. Each trip label occurs on a prefix and a continuation at $v$, giving $\wl=n$, but no route uses any trip in two runs, giving $\wr=0$. The segments may have arbitrarily many stops.

These bounds refer to the labels as given. Splitting a persistent resource into one label per session changes the charge, because $f(r_1+r_2)\le f(r_1)+f(r_2)$ is in general strict; it is legitimate only if the contract itself resets.

\subsection{Plan growth}

The following elementary reformulation of $L_f>0$ is used in Section~\ref{sec:hard}. Let $T^*_f(m)=\min\{T: J_f(T)\ge m\}$, with $T^*_f(m)=\infty$ if no such $T$ exists.

\begin{lemma}\label{lem:growth}
$L_f>0$ if and only if $T^*_f(m)$ is bounded by a polynomial in~$m$.
\end{lemma}
\begin{proof}
If $J_f(T)\ge T^\delta$ for $T\ge T_0$, then $T^*_f(m)\le\max\{T_0,\lceil m^{1/\delta}\rceil\}$. Conversely, if $T^*_f(m)\le Cm^{h}$ for all $m\ge m_0$, then for large $T$ and $m=\lfloor (T/C)^{1/h}\rfloor$ we get $J_f(T)\ge m\ge T^{1/(2h)}$, since $J_f$ is nondecreasing.
\end{proof}

\section{Exact compression by charging plans}\label{sec:alg}

\subsection{Supports and delayed commitment}

Fix a label $c$ with tariff $f_c$, marginals $\Delta_{c,j}$ and horizon~$T$. For $1\le j\le T$ the \emph{support} at $j$ is the affine function
\[
  \pi_{c,j}(r)=\beta_{c,j}+\Delta_{c,j}\,r,\qquad \beta_{c,j}=f_c(j)-j\Delta_{c,j}\ \ge 0 .
\]
Because the marginals are nonincreasing, $\pi_{c,j}(r)\ge f_c(r)$ for all $r\ge0$, with equality at $r=j$ and $r=j-1$. Supports with the same rate coincide, so there are $J_c(T):=J_{f_c}(T)$ distinct supports, and $f_c(r)=\min_j\pi_{c,j}(r)$ for $1\le r\le T$. The support at $1$ is the \emph{default} $\pi_{c,1}(r)=f_c(1)\,r$. We regard a support as a \emph{plan} with set-up fee $\beta_{c,j}$ and rate $\Delta_{c,j}$. More generally, a \emph{menu} for $c$ is any finite set of plans $\beta+\lambda r$ with $\beta\ge0$ and $0\le\lambda\le f_c(1)$ that contains the default, dominates $f_c$ on $\{0,\dots,T\}$ and attains $f_c(r)$ for every $1\le r\le T$. The canonical menu of supports need not be the smallest one: for a cap $f_c(r)=\min\{a_cr,b_c\}$ the two plans $a_cr$ and $b_c$ form a menu, while there are three distinct marginals when $b_c/a_c$ is not an integer.

\begin{lemma}[Delayed commitment]\label{lem:delayed}
Let a label pay as follows. At each new run it pays the default rate $f_c(1)$, until at the start of some run it \emph{commits} to a non-default plan $(\beta,\lambda)$ of its menu; at that run it pays $\beta+\lambda$, and at each later run it pays~$\lambda$. Then every such strategy pays at least $f_c(r)$ over $r$ runs, and for every $r$ some strategy pays exactly $f_c(r)$.
\end{lemma}
\begin{proof}
Never committing costs $f_c(1)r\ge f_c(r)$. Committing to $(\beta,\lambda)$ after $u\ge0$ default runs costs
\[
  f_c(1)\,u+\beta+\lambda(r-u)=\beta+\lambda r+\bigl(f_c(1)-\lambda\bigr)u\ \ge\ \beta+\lambda r\ \ge\ f_c(r).
\]
For exactness, commit at the first run to a plan attaining $f_c(r)$, or never commit if the default attains it.
\end{proof}

The state of a label under such a strategy is either one of the non-default plans or ``default''. A label that has not been used and a label that has been used at the default rate are in the same state, which is what makes the state space independent of run counts.

\subsection{The algorithm}

For per-label tariffs let $J=\max_cJ_c(T)$ for canonical menus, or the largest size of the supplied menus after deleting redundant plans. For a supplied menu, retain the default and one attaining plan for each integer $r=2,\dots,T$; this preserves exactness and ensures $J\le T$.

Let $D(T)$ be a common denominator of every retained plan fee and rate, including the default rates, and let $A(T)=D(T)\max_cf_c(1)$. For canonical supports it suffices to clear the tariff values $f_c(0),\dots,f_c(T)$, since their fees and rates are integer linear combinations of these values. For supplied menus the coefficients can require a finer grid. This distinction affects only part~(3) below; parts~(1) and~(2) use exact rational arithmetic without forming $D(T)$.

\begin{theorem}[Plan compression]\label{thm:plans}
For per-label tariffs the following hold.
\begin{enumerate}[label=\textup{(\arabic*)}]
\item A path of minimum charge can be computed in time $J^{\wr}\poly(N)$.
\item With integer lengths at most $W$, \textsc{Shortest Budgeted Path} can be solved in time $J^{\wr}\poly(N,W)$.
\item With arbitrary rational lengths, it can be solved in time $J^{\wr}\poly(N,A(T))$.
\end{enumerate}
\end{theorem}
\begin{proof}
Consider the process of Lemma~\ref{lem:projection} in which $h(c)\in\{\bot\}\cup\{\text{non-default plans of }c\}$, where $\bot$ means default. At an edge starting a run of $c$ with $h(c)=\bot$ the process either stays in $\bot$ and pays $f_c(1)$, or commits to a plan and pays its fee and rate; with $h(c)=(\beta,\lambda)$ it pays $\lambda$. Call the total payment along a path, for some choices, a \emph{certificate} of the path. By Lemma~\ref{lem:delayed}, applied to each label, every certificate of $P$ is at least $\Phi(P)$ and some certificate equals $\Phi(P)$. Hence the minimum certificate over all paths and choices equals $\min_P\Phi(P)$.

By Lemma~\ref{lem:projection} the same certificates are produced by the projected process, whose keys consist of a vertex~$v$, the last label, and a state for every label in $\Lr(v)$. There are at most $|V|(q+1)J^{\wr}$ keys, and each has at most $J$ transitions per outgoing edge.

(1) Process the keys in topological order of their vertices and keep for each key the minimum certificate of a prefix reaching it. A path performs at most $2T$ additions of fees and rates, each a rational number with polynomially many bits, so all numbers stay polynomially long and no common denominator is ever formed.

(2) Extend the key by the accumulated length in $\{0,\dots,TW\}$ and keep the minimum certificate for each extended key. The answer is the least length for which some key at $t$ has certificate at most~$B$.

(3) Extend the key instead by the integer $D(T)$ times the accumulated payment and keep the minimum length. Since payments are nonnegative and an exact certificate of a feasible path is at most $\Phi(P)\le T\max_cf_c(1)$, payments above $D(T)\cdot\min\{B,T\max_cf_c(1)\}$ are discarded, which leaves at most $TA(T)+1$ values.

In all three cases two records with the same extended key have the same transitions and future payments, so the dominated one can be discarded.
\end{proof}

\begin{corollary}\label{cor:fpt}
Let $\mathcal F$ be a class of per-label inputs for which there is a computable $\tau$ with $J\le T^{1/p}$ whenever $T\ge\tau(p)$. On $\mathcal F$, feasibility and \textsc{Shortest Budgeted Path} with polynomially bounded integer lengths are fixed-parameter tractable in~$\wr$. For rational lengths the same holds if in addition $A(T)$ is polynomially bounded in~$N$. In particular this applies to every fixed effectively subpolynomial tariff, and for rational lengths if it also lies on a polynomial grid.
\end{corollary}
\begin{proof}
Let $\wr\ge1$. If $T\ge\tau(\wr)$ then $J^{\wr}\le T$, and otherwise $J^{\wr}\le T^{\wr}<\tau(\wr)^{\wr}$.
\end{proof}

For example, $J(T)\le(\log(T+2))^{a}$ gives $J^{\wr}\le g_a(\wr)\,(T+2)$, and so does $J(T)\le\exp\bigl(a\sqrt{\log(T+2)}\bigr)$, because $a\wr\sqrt{L}\le a^2\wr^2/4+L$ for $L=\log(T+2)$.

\begin{example}[Huge denominators are harmless]\label{ex:tower}
Let $\Delta_j=2^{-2^{\lfloor\log_2j\rfloor}}$. Then $J_f(T)=1+\lfloor\log_2T\rfloor$, while $D_f(T)=2^{2^{\lfloor\log_2T\rfloor}}\ge2^{T/2}$. The numerators and denominators of all values up to $T$ each have at most $T+O(\log(T+1))$ bits, so Theorem~\ref{thm:plans}(1),(2) apply and the problem is fixed-parameter tractable, although no polynomial grid exists.
\end{example}

\begin{example}[Per-label caps]\label{ex:caps}
If every label has a cap tariff $f_c(r)=\min\{a_cr,b_c\}$ with its own binary-encoded coefficients, the two-plan menus give time $2^{\wr}\poly(N)$ for feasibility, whereas run counting needs up to $\min\{T,\lceil b_c/a_c\rceil\}+1$ states per label. For rational lengths the per-label magnitudes matter: already linear tariffs $f_c(r)=a_cr$ on a graph in which every label occurs once ($\wl=0$, $J=1$) encode the restricted shortest path problem, which is $\NP$-hard~\cite{GareyJ79}. This is why part~(3) involves $A(T)$.
\end{example}

\subsection{Fixed monetary resolution}

\begin{proposition}\label{prop:grid}
For a tariff $f$ the following are equivalent: \textup{(a)} all marginals lie in $\frac1{D_0}\Z$ for some integer~$D_0$; \textup{(b)} $f$ is eventually linear; \textup{(c)} $J_f$ is bounded. In this case \textsc{Shortest Budgeted Path} is fixed-parameter tractable in $\wr$ even for rational lengths.
\end{proposition}
\begin{proof}
If (a) holds, the marginals form a nonincreasing sequence in $\{0,\frac1{D_0},\dots,\Delta_1\}$, which has at most $D_0\Delta_1$ drops; this gives (c), and (c) implies that the marginals are eventually constant, which is (b). If (b) holds, there are finitely many distinct marginals and their common denominator serves as~$D_0$. Under (a) both $J_f$ and $D_f$ are bounded by constants, and Theorem~\ref{thm:plans}(3) applies.
\end{proof}

Thus every tariff quoted in a fixed currency unit is tractable, and so is every finite family of such tariffs from which labels choose. For an eventually linear fixed tariff one could equally clip run counts at the point of linearity; plans are needed for input-dependent menus such as Example~\ref{ex:caps} and for tariffs that never become linear. If per-label marginals lie on a grid $\frac1{D_0}\Z$ supplied with the input, then $J_c(T)\le\min\{T,D_0f_c(1)+1\}$, and feasibility is fixed-parameter tractable in $\wr+\max_cD_0f_c(1)$.

For a fixed tariff with $J_f(T)=\Theta(T^{\delta})$, $0<\delta<1$, Theorem~\ref{thm:plans} still improves the exponent of the baseline from $\wr$ to $\delta\wr$. The next section shows that it cannot be removed.

\section{Hardness for polynomial plan growth}\label{sec:hard}

Throughout this section $f$ is a fixed tariff in the common-tariff model.

\subsection{Jensen gaps and strict codes}

\begin{lemma}[Gap identity]\label{lem:gap}
Let $K\ge2$ and let $a_1,\dots,a_K\ge0$ be integers whose mean $m=\frac1K\sum_ia_i$ is an integer. Put $a_{\min}=\min_ia_i$ and $a_{\max}=\max_ia_i$. Then
\[
   f(m)-\frac1K\sum_{i=1}^Kf(a_i)\;=\;\sum_{t\ge1}d_t\,b_t,\qquad
   b_t=\frac1K\sum_{i=1}^K(a_i-t)_+-(m-t)_+ ,
\]
where $b_t=0$ unless $a_{\min}<t<a_{\max}$, and $b_t\ge\frac1K\min\{t-a_{\min},\,a_{\max}-t\}\ge\frac1K$ for $a_{\min}<t<a_{\max}$. In particular the left-hand side is nonnegative, and it vanishes if and
only if $f$ has no drop $t$ with $a_{\min}<t<a_{\max}$.
If $a_{\min}<a_{\max}$, it is at least
$\frac1K(\Delta_{a_{\min}+1}-\Delta_{a_{\max}})$;
if $a_{\min}=a_{\max}$, it is zero.
\end{lemma}
\begin{proof}
From $\Delta_j=\Delta_1-\sum_{t<j}d_t$ we get $f(r)=r\Delta_1-\sum_{t\ge1}d_t\,(r-t)_+$ for every integer $r\ge0$. The linear terms cancel in the left-hand side, which gives the identity. If $t\le a_{\min}$ then all positive parts equal their arguments and $b_t=0$; if $t\ge a_{\max}$ all terms vanish. Let $a_{\min}<t<a_{\max}$. If $t\ge m$ then $b_t=\frac1K\sum_i(a_i-t)_+\ge\frac1K(a_{\max}-t)$. If $t<m$ then $b_t=\frac1K\sum_i\bigl((a_i-t)_+-(a_i-t)\bigr)=\frac1K\sum_i(t-a_i)_+\ge\frac1K(t-a_{\min})$. If $a_{\min}=a_{\max}$, all coefficients $b_t$ are zero.
Otherwise
$\sum_{a_{\min}<t<a_{\max}}d_t
=\Delta_{a_{\min}+1}-\Delta_{a_{\max}}$,
which gives the stated lower bound.
\end{proof}

\begin{definition}\label{def:code}
Let $K\ge2$. A sequence of positive integers $r_1<\dots<r_z$ is a \emph{$K$-strict code} for $f$ if for every $x<z$ there is a drop $t$ of $f$ with $Kr_x<t<Kr_{x+1}$. Its \emph{size} is the number $r_z$.
\end{definition}

By Lemma~\ref{lem:gap} with $a_i=Kr_{x_i}$, a $K$-strict code satisfies, for all indices $x_1,\dots,x_K\in\{1,\dots,z\}$,
\begin{equation}\label{eq:EK}
  E_K(x_1,\dots,x_K):=f\Bigl(\sum_{i=1}^Kr_{x_i}\Bigr)-\frac1K\sum_{i=1}^Kf(Kr_{x_i})\ \ge0,\quad\text{with equality iff } x_1=\dots=x_K .
\end{equation}
We need codes whose \emph{size}, and not merely bit length, is polynomial, because run counts are realised by explicit paths. For instance, the tariff with $\Delta_j=1/(1+\lfloor\log_2j\rfloor)$ has the $2$-strict codes $r_x=4^x$, but none of polynomial size.

\begin{lemma}[Codes and plan growth]\label{lem:codes}
The following are equivalent.
\begin{enumerate}[label=\textup{(\alph*)}]
\item $L_f>0$.
\item There are constants $c,e$ such that for every $z$ there is a $2$-strict code of length $z$ and size at most $cz^{e}$.
\item There are constants $c,e$ such that for all $K\ge2$ and $z\ge1$ a $K$-strict code of length $z$ and size at most $c(Kz)^{e}$ exists and can be computed in time polynomial in $K$ and~$z$.
\end{enumerate}
\end{lemma}
\begin{proof}
(c) implies (b) trivially. Assume (b). The drops witnessing strictness for $x=1,\dots,z-1$ are distinct and smaller than $2r_z$, so $J_f(2r_z)\ge z$ and $T^*_f(z)\le 2cz^{e}$; Lemma~\ref{lem:growth} gives~(a).

Assume (a). By Lemma~\ref{lem:growth} there are constants with $T^*_f(m)\le Cm^{h}$ for $m\ge m_0$. Given $K$ and $z$, let $m=\max\{m_0,2Kz+1\}$ and compute marginals until $m$ distinct values have been seen; this happens at horizon $T=T^*_f(m)\le Cm^{h}$ and takes polynomial time. There are at least $2Kz$ drops below~$T$. Let $t_1<\dots<t_z$ be the $2K$-th, $4K$-th, \dots, $2Kz$-th of them, so that $t_{x+1}-t_x\ge2K$, and set $r_x=\lfloor t_x/K\rfloor+1$. Then $t_x<Kr_x\le t_x+K<t_{x+1}<Kr_{x+1}$, so the sequence is increasing and the drop $t_{x+1}$ lies strictly between $Kr_x$ and $Kr_{x+1}$. The size is $r_z\le T/K+1$.
\end{proof}

No bound on denominators enters this lemma. The algorithm in (c) need not know the constants: it simply scans until enough drops are found.

\subsection{Building blocks}

A label is \emph{fresh} if it is used on exactly one edge of the constructed graph. A fresh label is never live and contributes $f(1)$ to the charge of every path through its edge. A label is \emph{private} to a subpath if all its occurrences are on that subpath.

\begin{lemma}[Normalised block]\label{lem:block}
For integers $1\le r\le U$ and a label $c$ there is a word $\blk(c,r,U)$ of length $2U$ that contains $U$ occurrences of $c$, forming exactly $r$ runs, and $U$ fresh labels, the last of which is its final symbol.
\end{lemma}
\begin{proof}
Write $U-r+1$ copies of $c$, then $r-1$ times a fresh label followed by $c$, then $U-r+1$ fresh labels.
\end{proof}

Whenever such a block is preceded by a symbol other than $c$, it gives $c$ exactly $r$ new runs, and the other labels of the block are charged $Uf(1)$ regardless of~$r$.

\begin{lemma}[Selection gadget]\label{lem:select}
Let $W_1,\dots,W_n$, with $n\ge1$, be label words of common length $\ell\ge1$, and put $d=\lceil\log_2n\rceil$. There is a two-terminal series-parallel DAG with terminals $a,b$, of size $O(n\ell)$, such that every $a$--$b$ path spells $d$ fresh labels, some $W_i$, and $d$ fresh labels, and every $W_i$ is spelled by some path. The terminals have degree at most two and all other vertices have degree at most three.
\end{lemma}
\begin{proof}
Repeat words to obtain a list of exactly $2^{d}$ words. For one word take a path spelling it. For a list of $2^{j}$ words, $j\ge1$, take the gadgets of its two halves, new terminals $a,b$, a fresh-labelled edge from $a$ to the first terminal of each half and a fresh-labelled edge from the second terminal of each half to~$b$. Inner terminals receive one edge on the outside and have at most two on the inside.
\end{proof}

A \emph{chain} of selection gadgets joins the second terminal of each gadget to the first terminal of the next by a fresh-labelled \emph{bridge}. If the words within each gadget have equal length, then all source-to-target paths of the chain have the same number of edges, so the DAG is layered and hence bipartite. It is simple, series-parallel, hence planar and of treewidth at most two~\cite{Bodlaender93}, and its maximum degree is three. All paths contain the same number $g$ of fresh selector and bridge edges.

\subsection{The reduction}

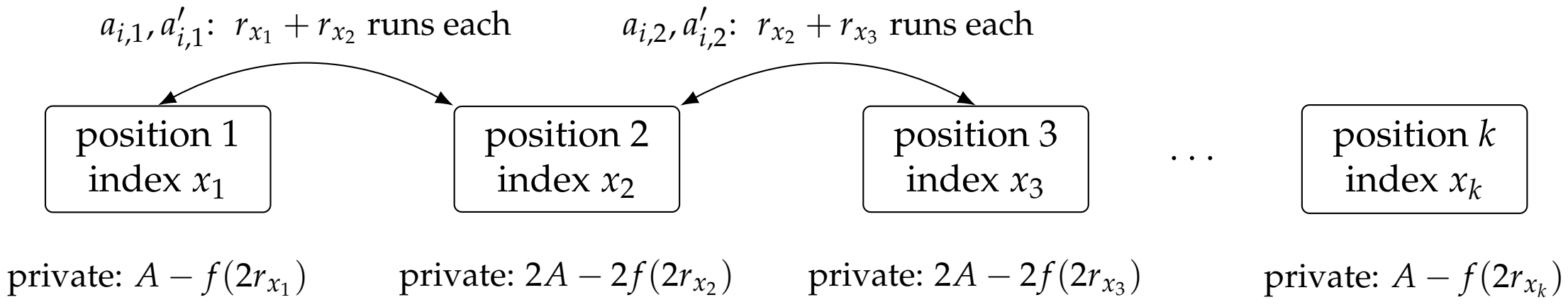
\begin{figure}[t]
\centering
\begin{tikzpicture}[>=Latex,font=\small,
  mod/.style={draw,rounded corners=2pt,minimum height=9mm,minimum width=21mm,align=center},
  chk/.style={font=\footnotesize,align=center}]
\node[mod] (m1) {position $1$\\ index $x_1$};
\node[mod,right=17mm of m1] (m2) {position $2$\\ index $x_2$};
\node[mod,right=17mm of m2] (m3) {position $3$\\ index $x_3$};
\node[right=6mm of m3] (dots) {$\cdots$};
\node[mod,right=6mm of dots] (mk) {position $k$\\ index $x_k$};
\draw[<->] (m1.north) to[bend left=28] node[chk,above]{$a_{i,1},a'_{i,1}$: $\;r_{x_1}+r_{x_2}$ runs each} (m2.north west);
\draw[<->] (m2.north east) to[bend left=28] node[chk,above]{$a_{i,2},a'_{i,2}$: $\;r_{x_2}+r_{x_3}$ runs each} (m3.north);
\node[chk,below=3mm of m1] {private: $A-f(2r_{x_1})$};
\node[chk,below=3mm of m2] {private: $2A-2f(2r_{x_2})$};
\node[chk,below=3mm of m3] {private: $2A-2f(2r_{x_3})$};
\node[chk,below=3mm of mk] {private: $A-f(2r_{x_k})$};
\end{tikzpicture}
\caption{The $k$ modules of one colour~$i$, in the order in which a path meets them. Consecutive modules share two check labels. The check between positions $m$ and $m+1$ contributes $2f(r_{x_m}+r_{x_{m+1}})-f(2r_{x_m})-f(2r_{x_{m+1}})\ge0$ above a constant, with equality exactly if $x_m=x_{m+1}$. At any vertex at most two check labels of the colour are live.}
\label{fig:chain}
\end{figure}

\textsc{Multicolored Clique} asks, for a graph $H$ whose vertex set is partitioned into colour classes $V_1,\dots,V_k$, whether $H$ has a clique with one vertex in each class. It is $\W1$-hard parameterized by~$k$~\cite{FellowsHRV09}. We assume $k\ge2$, write $V_i=\{v_{i,1},\dots,v_{i,z_i}\}$, and put $z=\max\{2,z_1,\dots,z_k\}$. Let $E_{ij}$ be the set of edges between $V_i$ and $V_j$; if some $V_i$ or $E_{ij}$ is empty the answer is no.

\begin{samepage}
\begin{theorem}\label{thm:hard}
Let $f$ be a fixed tariff with $L_f>0$. There is a polynomial-time algorithm that maps an instance $(H,k)$ of \textsc{Multicolored Clique} to an equivalent instance of \textsc{Concave Run Budget} for $f$ with $\wr\le\wl\le2k+1$ whose graph is a simple planar bipartite DAG of maximum degree three and treewidth at most two in which all paths have the same number of edges. Consequently \textsc{Concave Run Budget} for $f$ is $\NP$-complete and $\W1$-hard parameterized by $\wr$ or by $\wl$ on this class of graphs, and unless ETH fails it has no algorithm with running time $g(\wr)N^{o(\wr)}$ for any computable~$g$.
\end{theorem}
\end{samepage}

\begin{proof}
Since $L_f>0$, we have $f(1)>0$: otherwise all marginals would vanish.
If some colour class $V_i$ or some set $E_{ij}$ is empty, output the
one-edge $s$--$t$ graph with budget zero. Its unique path has charge
$f(1)>0$, and its live width is zero. We may therefore assume that
all these sets are nonempty. Code entries not used as vertex indices are allowed in the
compensating private blocks.

\emph{Code.} By Lemma~\ref{lem:codes} we compute a $2$-strict code $r_1<\dots<r_z$ of size $U:=r_z$ polynomial in~$z$. For indices $x,y$ let
\[
  \gamma(x,y)=2f(r_x+r_y)-f(2r_x)-f(2r_y)=2E_2(x,y),
\]
so $\gamma(x,y)\ge0$ with equality exactly if $x=y$, by \eqref{eq:EK}.

\emph{Gadgets and positions.} The graph is a chain of selection gadgets: a vertex gadget $\mathcal V_i$ for each colour~$i$, followed by an edge gadget $\mathcal E_{ij}$ for each pair $i<j$. Colour $i$ is \emph{involved} in $\mathcal V_i$ and in the $k-1$ gadgets $\mathcal E_{ij}$, $j\ne i$; we number these $k$ gadgets $1,\dots,k$ in chain order and call the number the \emph{position} of the gadget for colour~$i$. For every colour $i$ and every $m\in\{1,\dots,k-1\}$ there are two \emph{check labels} $a_{i,m}$ and $a'_{i,m}$.

\emph{Parts and modules.} For a colour $i$, a check $m$ and an index $x$, the \emph{part} $Q_{i,m}(x)$ is the word
\[
  \blk(a_{i,m},r_x,U)\;\blk(a'_{i,m},r_x,U)\;\blk(p_1,\rho_1,2U)\cdots\blk(p_z,\rho_z,2U),\qquad
  \rho_y=\begin{cases}2r_y&y\ne x,\\ 1& y=x,\end{cases}
\]
where $p_1,\dots,p_z$ are new labels private to their blocks. It has length $4U(z+1)$. Apart from the two check labels, its labels are charged
\begin{equation}\label{eq:part}
  2Uf(1)+\sum_{y\ne x}f(2r_y)+f(1)+2Uzf(1)\;=\;A-f(2r_x),
\end{equation}
where $A:=\sum_{y=1}^zf(2r_y)+\bigl(2U(z+1)+1\bigr)f(1)$ does not depend on~$x$.
The \emph{module} $M_{i,m}(x)$ for position $m$ is $Q_{i,m-1}(x)\,Q_{i,m}(x)$, where $Q_{i,0}$ and $Q_{i,k}$ are empty. Its length depends only on $m$.

\emph{Payloads.} The words of $\mathcal V_i$ are the modules $M_{i,m}(x)$ for $x=1,\dots,z_i$, where $m$ is the position of $\mathcal V_i$ for colour~$i$. The words of $\mathcal E_{ij}$ are $M_{i,m}(x)\,M_{j,m'}(y)$ for all edges $v_{i,x}v_{j,y}\in E_{ij}$, with $m,m'$ the positions of $\mathcal E_{ij}$ for $i$ and~$j$. Private and fresh labels are new in every word, including repeated copies. The budget is
\[
  B_0=2k(k-1)A+g\,f(1),
\]
where $g$ is the number of selector and bridge edges on a path.

\emph{Charge of a path.} A path selects one word in every gadget and thus, for each colour $i$, an index $x_{i,m}$ at each position $m$. Every block either starts the path or is preceded by a fresh label, possibly the final label of another block, so runs never merge across blocks. The label $a_{i,m}$ occurs only in the parts $Q_{i,m}$ at positions $m$ and $m+1$ and therefore has $r_{x_{i,m}}+r_{x_{i,m+1}}$ runs; the same holds for $a'_{i,m}$. A path traverses $2k(k-1)$ parts, two for each check, and by \eqref{eq:part}
\[
  \Phi(P)=B_0+\sum_{i=1}^k\sum_{m=1}^{k-1}\gamma\bigl(x_{i,m},x_{i,m+1}\bigr).
\]
Hence $\Phi(P)\ge B_0$, with equality if and only if $x_{i,1}=\dots=x_{i,k}$ for every~$i$, that is, if and only if every selected edge in $E_{ij}$ joins the selected vertices of $V_i$ and $V_j$. Such a path exists exactly if $H$ has a multicoloured clique.

\emph{Width.}
Write $\mathcal G_1,\dots,\mathcal G_h$ for the gadgets in chain
order. For a colour $i$, let
$j_{i,1}<\dots<j_{i,k}$ be the indices of its involved gadgets,
and put $C_{i,m}=\{a_{i,m},a'_{i,m}\}$ for $1\le m<k$ and
$C_{i,0}=C_{i,k}=\varnothing$.
Labels in $C_{i,m}$ occur only in
$\mathcal G_{j_{i,m}}$ and $\mathcal G_{j_{i,m+1}}$.

First consider a vertex in a gadget that does not involve $i$.
If exactly $m$ involved gadgets precede it, all check labels other
than those in $C_{i,m}$ have both occurrence gadgets on the same
side, so only $C_{i,m}$ can be live there.
The same argument applies at both endpoints of a bridge: take $m$
to be the number of involved gadgets up to and including the
gadget before that bridge. The bridge itself is a single fresh
edge and has no internal vertices.

Now let $v$ lie in the involved gadget
$\mathcal G_{j_{i,m}}$. Only the labels in
$C_{i,m-1}\cup C_{i,m}$ can be live at $v$.
View the selection gadget as a splitting tree, internally disjoint
payload paths, and a merging tree.
At a vertex of the splitting tree, including a payload start,
every prefix traverses only selector edges within this gadget.
Since labels in $C_{i,m}$ first occur in this gadget,
$C_{i,m}\cap\Pi(v)=\varnothing$.
At a vertex of the merging tree, including a payload end,
every continuation traverses only selector edges within this
gadget. Since labels in $C_{i,m-1}$ last occur in this gadget,
$C_{i,m-1}\cap\Gamma(v)=\varnothing$.
These statements also cover the gadget terminals and the case of
a single payload, for which the selector trees have depth zero.

If $v$ is strictly inside a payload path, every prefix and
continuation through $v$ uses that same payload path inside the
gadget. In this word all occurrences of $C_{i,m-1}$ precede all
occurrences of $C_{i,m}$, because the module is
$Q_{i,m-1}(x)Q_{i,m}(x)$; any module of another colour is wholly
before or after it. If the payload prefix up to $v$ contains no
edge labelled from $C_{i,m}$, then
$C_{i,m}\cap\Pi(v)=\varnothing$. Otherwise all occurrences of
$C_{i,m-1}$ in this payload precede $v$, and this check has no
later occurrence gadget, so
$C_{i,m-1}\cap\Gamma(v)=\varnothing$.
Thus at most one of the two check pairs can contribute live
labels, and at most two check labels of colour $i$ are live at
every vertex.

Fresh labels are never live. A private label is confined to one
block on one payload copy. If it is live at $v$, a prefix and a
continuation witnessing this concatenate to a path through two
occurrences of that label, forcing $v$ to be an internal vertex
of its block. Distinct private blocks have disjoint interiors,
so at most one private label is live at a vertex. Consequently
$\wl\le2k+1$, and $\wr\le\wl$ by definition.

\emph{Size and graph class.}
Every payload has at most $16U(z+1)$ symbols. Each gadget has at
most $z^2$ candidate words before padding and fewer than $2z^2$
payload copies after padding. There are $k+\binom k2$ gadgets,
so the numbers of vertices and edges are
$O(k^2z^2U(z+1))$. Since $U$ is polynomial in $z$, this is
polynomial in $k$ and $z$, with an exponent independent of $k$.
The values of $f$ up to $2U$ have polynomial bit length and can
be computed in polynomial time. The budget is a sum of
polynomially many such rational values with polynomially bounded
integer coefficients, so it too has polynomial bit length and
is exactly computable in polynomial time. Encoding vertex and
label identifiers therefore gives $N\le(kz)^{O(1)}$.
The graph properties were established after
Lemma~\ref{lem:select}.

\emph{Consequences.} The reduction is polynomial and maps $k$ to a width of at most $2k+1$; this proves $\NP$-hardness and $\W1$-hardness. Membership in $\NP$ holds because the charge of a given path is computed in polynomial time. Under ETH, \textsc{Multicolored Clique} has no
$g(k)n^{o(k)}$ algorithm for computable $g$
\cite[Corollary~14.23]{CyganEtAl15} (see also~\cite{ChenHKX06}).
Suppose feasibility had running time $g(p)N^{e(p)}$ for
$p=\wr$, with $e(p)=o(p)$.
Put
$G(p)=\max_{0\le q\le p}\max\{1,g(q)\}$ and
$E(p)=\max_{0\le q\le p}\max\{0,e(q)\}$.
Then $G$ is computable and $E(p)=o(p)$: for any $\eta>0$,
the finitely many values before $e(q)\le\eta q$ contribute
only a fixed constant to the maximum.
For a constant $C$ the constructed instance has
$N\le(kz)^C$ and $p\le2k+1$, so the target running time is at
most
$G(2k+1)(kz)^{C E(2k+1)}$.
Here $C E(2k+1)=o(k)$, and the factor depending only on $k$
is bounded by a computable function, since
$E(2k+1)=O(k)$.
Including the polynomial reduction time therefore gives a
$g'(k)z^{o(k)}$ algorithm with computable $g'$, contradicting
the cited lower bound because $z\le\max\{2,n\}$.
\end{proof}

Up to constant factors in the exponent the lower bound matches the count-state algorithm of Proposition~\ref{prop:baseline}, whose running time is $N^{O(\wr)}$.

\begin{remark}[Separation gap]\label{rem:gap}
In the constructed instances every path has charge exactly $B_0$ or at least $B_0+\gamma_{\min}$, where by Lemma~\ref{lem:gap}
\[
  \gamma_{\min}=\min_{x\ne y}\gamma(x,y)\ \ge\ \min_{x<z}\bigl(\Delta_{2r_x+1}-\Delta_{2r_{x+1}}\bigr)>0 .
\]
If $f$ lies on a polynomial grid, all charges and $B_0$ are multiples of $1/D_f(2U)$, so $\gamma_{\min}\ge1/D_f(2U)$ is inverse-polynomial. In general $\gamma_{\min}$ may be exponentially small, as for geometric marginals.
\end{remark}

\begin{remark}\label{rem:onlyedges}
The construction uses private labels generously. The number of labels occurring more than once is polynomial in $k$ and $z$, so the theorem says nothing about the parameters $q$ and $\qrep$, which are treated in Section~\ref{sec:few}.
\end{remark}

\section{Width \texorpdfstring{$k+1$}{k+1} on polynomial grids}\label{sec:residue}

The reduction of Section~\ref{sec:hard} keeps two check labels per colour alive. With a single shared label $c_i$ per colour, receiving $r_x$ runs in each of the $k$ gadgets that involve the colour, the charge of $c_i$ is $f(\sum_mr_{x_{i,m}})$, and \eqref{eq:EK} with $K=k$ shows that consistency is detected if each of these gadgets additionally pays a constant minus $f(kr_x)/k$. This quantity is not a sum of tariff values. We show that on a polynomial grid it can nevertheless be paid by positive charges, after a suitable choice of the code. The result is the following refinement.

\begin{theorem}\label{thm:residue}
Let $f$ be a fixed tariff on a polynomial grid with $L_f>0$. Then the statement of Theorem~\ref{thm:hard} holds with $\wr\le\wl\le k+1$ in place of $2k+1$.
\end{theorem}

\begin{lemma}[Positive padding]\label{lem:pad}
Let $f(1)>0$, let $R\ge1$, let $D$ be a positive integer with $f(0),\dots,f(R)\in\frac1D\Z$, and put
\[
  v_r=D\bigl(f(r)+(r-1)f(1)\bigr),\quad d=\gcd(v_1,\dots,v_R),\quad u_r=v_r/d,\quad a=\max_ru_r,
\]
where $r$ ranges over $1,\dots,R$.
\begin{enumerate}[label=\textup{(\roman*)}]
\item $Df(r)\in d\Z$ for all $0\le r\le R$.
\item Every integer $n\ge a^2$ is a sum of at most $n$ numbers from $\{u_1,\dots,u_R\}$, and such a representation can be computed in time polynomial in $R$, $a$ and $n$.
\item For every integer $n\ge a^2$ and every integer $\ell\ge(2R-1)n$, one can construct in time polynomial in $R,n,\ell$ a word of length exactly $\ell$ and charge exactly $dn/D$ in which no label has more than $R$ runs, all labels are new, and at most one label is live at any position.
\end{enumerate}
\end{lemma}
\begin{proof}
(i) $Df(r)=v_r-(r-1)v_1$.

(ii) The $u_r$ are positive integers with greatest common divisor one. Let $\mu=\min_ru_r$ and consider the digraph on $\Z/\mu\Z$ with an arc of weight $u_r$ from $x$ to $x+u_r$ for all $x$ and~$r$. The residues reachable from $0$ form a subsemigroup of a finite group, hence a subgroup, which is the whole group because the $u_r$ are coprime. A shortest path from $0$ to the residue of $n$ uses at most $\mu-1$ arcs and has weight at most $(\mu-1)a<a^2\le n$; adding copies of $\mu$ gives $n$. Shortest paths are found by Dijkstra's algorithm.

(iii) For a term $u_r$ write a new label $c$ followed by $r-1$ times a fresh label and $c$. This word has charge $f(r)+(r-1)f(1)=du_r/D$. Concatenate the words of all terms of a representation of~$n$, with new labels in each; consecutive words begin and end with different labels. The only labels with more than one occurrence are the labels $c$, each confined to its word. There are at most $n$ terms and each block has length $2r-1\le2R-1$, so the concatenation has length at most $(2R-1)n$. Repeating its first symbol reaches any prescribed $\ell\ge(2R-1)n$ without adding a run.
\end{proof}

\begin{lemma}[Residue-selected modules]\label{lem:resmod}
Let $f$ lie on a polynomial grid and $L_f>0$. For all $K\ge2$ and $z\ge1$ one can compute, in time polynomial in $K$ and~$z$, a $K$-strict code $r_1<\dots<r_z$ of size $U$, a rational $C_0>0$ and an integer $\ell$ such that for every $x$ and every label $c$ there is a word $R_c(x)$ of length $\ell$ that consists of $\blk(c,r_x,U)$ followed by a word in new labels, has at most one live label other than $c$ at any position, and whose labels other than $c$ are charged exactly
\[
   C_0-\frac{f(Kr_x)}{K}.
\]
All charges of labels with at most $KU$ runs are multiples of $1/D$ for an integer $D$ bounded by a polynomial in $K$ and $z$, and $KC_0\in\frac1D\Z$.
\end{lemma}
\begin{proof}
By Lemma~\ref{lem:codes}(c) compute a $K$-strict code $t_1<\dots<t_{Kz}$ of length $Kz$; let $U=t_{Kz}$, $R=KU$ and $D=D_f(R)$, which are polynomially bounded, and define $d$, $u_r$ and $a$ as in Lemma~\ref{lem:pad}. As in the proof of Theorem~\ref{thm:hard}, $f(1)>0$. By Lemma~\ref{lem:pad}(i) the numbers $\nu_y=Df(Kt_y)/d$ are nonnegative integers. Some residue class $b\in\{0,\dots,K-1\}$ modulo $K$ contains at least $z$ of them; keep $z$ such candidates and call them $r_1<\dots<r_z$. A subsequence of a $K$-strict code is $K$-strict. The integers
\[
   m_x=\frac1K\Bigl(\frac{Df(Kr_x)}{d}-b\Bigr)\ \ge0,\qquad h=a^2+\max_xm_x,\qquad C=\frac dD\Bigl(h+\frac bK\Bigr)
\]
satisfy
\[
   C-\frac{f(Kr_x)}{K}=\frac dD\,(h-m_x),\qquad h-m_x\ge a^2 .
\]
Put $L=(2R-1)h$. Since $a^2\le h-m_x\le h$, Lemma~\ref{lem:pad}(iii) constructs, for every $x$, a padding word of length exactly $L$ and charge $\frac dD(h-m_x)$. Let $R_c(x)$ be $\blk(c,r_x,U)$ followed by this word, set $\ell=2U+L$, and set $C_0=Uf(1)+C$. The block ends with a fresh label, so no runs merge. Since $a\le 2RDf(1)$ and $m_x\le Df(R)\le DRf(1)$, the numbers $h$ and $\ell$ are polynomial. Finally $KC=\frac dD(Kh+b)\in\frac1D\Z$.
\end{proof}

The role of the residue class is to make $f(Kr_x)/K$ congruent, for all retained $x$, to one fixed element modulo the lattice $\frac dD\Z$ of realisable charges. The pool of $Kz$ candidates and the numbers $D$ and $d$ are fixed before the selection, so there is no circular dependence between the code and its lattice.

\begin{proof}[Proof of Theorem~\ref{thm:residue}]
Apply Lemma~\ref{lem:resmod} with $K=k$. Use the chain of gadgets of Theorem~\ref{thm:hard}, one shared label $c_i$ per colour, the words $R_{c_i}(x)$ in $\mathcal V_i$ and the words $R_{c_i}(x)R_{c_j}(y)$ in $\mathcal E_{ij}$, with new private labels in every word. Each colour is involved in exactly $K=k$ gadgets, so a path selecting indices $x_{i,1},\dots,x_{i,k}$ for colour $i$ has charge
\[
  \Phi(P)=k^2C_0+gf(1)+\sum_{i=1}^kE_k\bigl(x_{i,1},\dots,x_{i,k}\bigr),
\]
because $c_i$ is charged $f(\sum_mr_{x_{i,m}})$ and the other labels of its $k$ words are charged $kC_0-\frac1k\sum_mf(kr_{x_{i,m}})$. By \eqref{eq:EK} the budget $B_0=k^2C_0+gf(1)$ is met exactly by the paths that encode a multicoloured clique. At most the $k$ shared labels and one private label are live at a vertex. The remaining assertions are proved as in Theorem~\ref{thm:hard}. Every label has at most $kU$ runs, so all charges and $B_0$ lie in $\frac1D\Z$, and inconsistent paths exceed the budget by at least $1/D$.
\end{proof}

\section{Concrete tariffs, close tariffs and separation}\label{sec:examples}

\subsection{Hard tariffs}

A tariff is \emph{strictly concave} if every $t\ge1$ is a drop; then $J_f(T)=T$, $L_f=1$, and $r_x=x$ is a $K$-strict code for every $K\ge2$. Theorem~\ref{thm:hard} therefore applies to the harmonic tariff $H(r)=\sum_{j\le r}1/j$, to the reciprocal-power tariffs with $\Delta_j=j^{-p}$ for a fixed integer $p\ge1$, and to the bounded geometric tariff with $\Delta_j=2^{1-j}$. None of these lies on a polynomial grid. The following tariffs do.

Let $u(j)=2^{\lfloor\log_2j\rfloor}$ and define $F$ and, for a fixed integer $\alpha\ge1$, $G_\alpha$ by
\begin{equation}\label{eq:F}
  \Delta^F_j=\frac{3u(j)-j}{2u(j)^2},\qquad \Delta^{G_\alpha}_j=\Delta^F_{\lceil j^{1/\alpha}\rceil}.
\end{equation}

\begin{proposition}\label{prop:FG}
\begin{enumerate}[label=\textup{(\roman*)}]
\item $F$ is a strictly concave tariff with $d_t=1/(2u(t)^2)$ for all $t$, $D_F(T)\le2T^2$ and $F(r)=\Theta(\log(r+1))$.
\item $G_\alpha$ is a tariff whose drops are exactly the perfect $\alpha$-th powers; $J_{G_\alpha}(T)=\lceil T^{1/\alpha}\rceil$, $D_{G_\alpha}(T)\le8T^{2/\alpha}$, and $G_\alpha(r)=\Theta(r^{1-1/\alpha})$ for $\alpha\ge2$.
\end{enumerate}
\end{proposition}
\begin{proof}
(i) Within a dyadic range $u\le j<2u$ consecutive marginals differ by $1/(2u^2)$. The last marginal of the range is $(u+1)/(2u^2)$ and the first of the next range is $1/(2u)$, which again differ by $1/(2u^2)$. All marginals up to $T$ have denominators dividing $2u(T)^2\le2T^2$, and the marginals of a dyadic range sum to $\frac34+\frac1{4u}$. (ii) The index map $j\mapsto\lceil j^{1/\alpha}\rceil$ is nondecreasing and increases exactly after perfect $\alpha$-th powers; the remaining claims follow from (i), $\lceil T^{1/\alpha}\rceil\le2T^{1/\alpha}$ and $\Delta^F_j=\Theta(1/j)$.
\end{proof}

\begin{corollary}\label{cor:examples}
\textsc{Concave Run Budget} is $\NP$-complete and $\W1$-hard in $\wr$ on the graph class of Theorem~\ref{thm:hard}, with the ETH bound stated there, for every strictly concave tariff with polynomial-time evaluation, in particular for $H$, for reciprocal powers and for geometric marginals, and for $F$ and every $G_\alpha$. For $F$ and $G_\alpha$ the instances can be taken of width at most $k+1$, and \textsc{Shortest Budgeted Path} with rational lengths is in $\XP$ for these two families. For every $\delta>0$ there is a hard tariff with $J_f(T)=O(T^{\delta})$.
\end{corollary}
\begin{proof}
Theorems~\ref{thm:hard} and~\ref{thm:residue} and Proposition~\ref{prop:FG}. For the $\XP$ statement extend the keys of Proposition~\ref{prop:baseline} by the scaled charge as in Theorem~\ref{thm:plans}(3). The last claim is $G_\alpha$ with $\alpha>1/\delta$.
\end{proof}

\subsection{Arbitrarily close tariffs of different complexity}

For a fixed integer $M\ge1$ let $F_M$ be the tariff with marginals
\[
   \Delta^{F_M}_j=\frac{\lceil M\,u(j)\,\Delta^F_j\rceil}{M\,u(j)} .
\]

\begin{proposition}\label{prop:twins}
For every $M\ge1$, $F_M$ is a tariff with $F\le F_M\le(1+\frac2M)F$, $J_{F_M}(T)\le\lceil M/2\rceil(1+\log_2T)$ and $D_{F_M}(T)\le MT$. Consequently \textsc{Shortest Budgeted Path} for $F_M$ is fixed-parameter tractable in $\wr$, even with rational lengths, whereas feasibility for $F$ is $\W1$-hard.
\end{proposition}
\begin{proof}
Within a dyadic range with $u=u(j)$ the numbers $Mu\Delta^F_j=M(3u-j)/(2u)$ decrease and lie in $(M/2,M]$, so their ceilings are nonincreasing and take at most $\lceil M/2\rceil$ values. The first marginal of the next range is exactly $1/(2u)$, which is at most the last rounded marginal of the current one. Hence $F_M$ is a tariff with the stated plan count, and its denominators divide $M u(T)\le MT$. Moreover $\Delta^F_j\le\Delta^{F_M}_j<\Delta^F_j+\frac1{Mu}\le(1+\frac2M)\Delta^F_j$, because $\Delta^F_j>\frac1{2u}$. The algorithmic claim is Corollary~\ref{cor:fpt}, and hardness of $F$ is Corollary~\ref{cor:examples}.
\end{proof}

Since $F\le F_M\le(1+\frac2M)F$, a shortest path for $F_M$ with budget $(1+\frac2M)B$ is no longer than the optimum for $F$ with budget $B$ and exceeds the budget by a factor of at most $1+\frac2M$ with respect to~$F$. This is the principle behind the budget-relaxing scheme of Section~\ref{sec:relax}, and Corollary~\ref{cor:accuracy} shows that the price $g(\wr,M)$ of this relaxation cannot be made polynomial in $M$.

\subsection{Separation}

\begin{definition}\label{def:sep}
A fixed tariff $f$ with $L_f>0$ is \emph{polynomially separated} if there are constants $C\ge1$, $a,b\ge0$, and an algorithm which, for every integer $z\ge2$, runs in time at most $Cz^a$ and outputs a $2$-strict code $r_1<\dots<r_z$ with $r_z\le Cz^a$ and
\[
 \gamma_{\min}=\min_{1\le x<y\le z}\bigl(2f(r_x+r_y)-f(2r_x)-f(2r_y)\bigr)
       \ge C^{-1}z^{-b}.
\]
The constants and algorithm may depend on $f$, but not on $z$, the clique parameter, or the graph instance.
\end{definition}

\begin{lemma}\label{lem:sep}
A tariff $f$ with $L_f>0$ is polynomially separated if it lies on a polynomial grid, or if there are constants $c>0$ and $p$ with $d_t\ge ct^{-p}$ for all $t\ge1$.
\end{lemma}
\begin{proof}
On a polynomial grid, $\gamma(x,y)$ is a positive multiple of $1/D_f(2r_z)$. In the second case $f$ is strictly concave, $r_x=x$ is a $2$-strict code, and by Remark~\ref{rem:gap}, $\gamma_{\min}\ge\min_{x<z}d_{2x+1}\ge c(2z)^{-p}$.
\end{proof}

Thus $F$, $G_\alpha$, $H$ and the reciprocal powers are polynomially separated, the latter two with $d_t\ge p\,(t+1)^{-p-1}$. The class is stable under perturbations that preserve a polynomial lower bound on the decrements; for example $\Delta_j=1/j+\sum_{i=2}^{n}a_i/j^{i}$ with rational $a_i\ge0$ has $d_t\ge1/(2t^2)$. Geometric marginals have $L_f=1$ but are not polynomially separated: there $\gamma(x,y)=2\,(2^{-r_x}-2^{-r_y})^2$, which is exponentially small in $z$ for some pair of any code of length~$z$.

\section{Budget relaxation}\label{sec:relax}

With rational lengths, Theorem~\ref{thm:plans}(3) depends on the numerical range $A(T)$ of the charges, and plan counts may be large. Both dependencies disappear if the budget may be exceeded slightly.

The construction combines piecewise-linear approximation of concave costs~\cite{MagnantiS04} with the scaling and rounding used for restricted shortest paths~\cite{Hassin92,LorenzR01,TsaggourisZ09}. The additional point is that the approximate plans retain the delayed-commitment property and can be stored only on the restart set.

\begin{lemma}[Value-spaced menus]\label{lem:menu}
Let $f_c(1)>0$ and $0<\eta\le1$. One can compute a set of at most $2+\log_{1+\eta}\bigl(f_c(T)/f_c(1)\bigr)=O(1+\eta^{-1}\log(T+1))$ supports of $f_c$, including the default, whose minimum $\hat f_c$ satisfies $f_c(r)\le\hat f_c(r)\le(1+\eta)f_c(r)$ for $1\le r\le T$.
\end{lemma}
\begin{proof}
Start with $x=1$. Given $x\le T$, let $t\le T$ be largest with $f_c(t)\le(1+\eta)f_c(x)$, add the support at $t$ and continue with $x=t+1$. For $x\le r\le t$ we have $f_c(r)\le\pi_{c,t}(r)\le\pi_{c,t}(t)=f_c(t)\le(1+\eta)f_c(x)\le(1+\eta)f_c(r)$. The values $f_c(x)$ at consecutive starting points grow by a factor larger than $1+\eta$ and lie between $f_c(1)$ and $f_c(T)\le Tf_c(1)$.
\end{proof}

\begin{theorem}[Budget-relaxing scheme]\label{thm:relax}
For per-label tariffs, rational lengths and $0<\eps\le1$, one can compute in time $J_{\mathrm{app}}^{\wr}\poly(N,1/\eps)$, where $J_{\mathrm{app}}=O(1+\eps^{-1}\log(T+1))$, either a path $P$ with
\[
   \ell(P)\le\OPT(B)\qquad\text{and}\qquad\Phi(P)\le(1+\eps)B,
\]
where $\OPT(B)$ is the least length of a path of charge at most $B$, or a correct statement that no path has charge at most~$B$. The running time is of the form $g(\wr,1/\eps)\poly(N)$.
\end{theorem}
\begin{proof}
If $B=0$, only labels with $f_c(1)=0$ can be used; their tariffs vanish identically, and a shortest path in the corresponding subgraph is the answer. Let $B>0$ and $\rho=\eps B/(4T)$. Labels with $f_c(1)=0$ have the zero tariff and a single plan. For every other label take the menu of Lemma~\ref{lem:menu} with $\eta=\eps/2$, and round all fees and rates up to multiples of $\rho$. Run the projected plan process of Theorem~\ref{thm:plans} with these rounded plans, extend the key by the accumulated rounded payment divided by $\rho$, an integer that is kept only up to $\lfloor(1+\eps)B/\rho\rfloor=O(T/\eps)$, and store the minimum length for each extended key.

The proof of Lemma~\ref{lem:delayed} applies verbatim to the plans of Lemma~\ref{lem:menu}: every strategy pays at least $\hat f_c(r)\ge f_c(r)$, and committing at the first run to the plan attaining $\hat f_c(r)$ pays exactly $\hat f_c(r)$. Rounding up only increases payments, so every path found has $\Phi(P)\le(1+\eps)B$. Conversely let $P^*$ be a path of length $\OPT(B)$ and charge at most~$B$. Its exact strategies for the $\hat f_c$ pay at most $(1+\eps/2)B$ before rounding, and involve at most $T$ rates and at most $T$ fees, so rounding adds at most $2T\rho=\eps B/2$. All prefixes of this trajectory survive, and the algorithm returns a path of length at most $\ell(P^*)$.

There are at most $|V|(q+1)J_{\mathrm{app}}^{\wr}\,O(T/\eps)$ extended keys. Finally $(\log(T+1)/\eps)^{\wr}\le\eps^{-\wr}(\wr/e)^{\wr}\,(T+1)$ for $\wr\ge1$, which gives the form $g(\wr,1/\eps)\poly(N)$.
\end{proof}

\begin{corollary}[Accuracy obstruction]\label{cor:accuracy}
Let $f$ be a fixed polynomially separated tariff. Unless $\FPT=\W1$, no algorithm achieves the guarantee of Theorem~\ref{thm:relax} in time $g(\wr)\poly(N,1/\eps)$, even on unit-length instances that have a feasible path and whose graphs are simple planar bipartite DAGs of maximum degree three and treewidth at most two.
\end{corollary}
\begin{proof}
First decide any source instance with an empty colour class $V_i$ or an empty edge set $E_{ij}$ as a no-instance, without constructing a fallback. For the remaining sources, take the instance of Theorem~\ref{thm:hard}, built from a polynomially separated code, and let $L_0$ be the common length of its paths. Here $B_0\ge f(1)>0$. Add a \emph{fallback} path of $3L_0$ edges from $s$ to $t$, all carrying one new label. Its charge is $f(1)\le B_0$, so the instance is feasible. The parities of $L_0$ and $3L_0$ agree, so the graph remains bipartite; it remains series-parallel and of maximum degree three, and the new label is live only on the fallback path. Let $\eps=\min\{1,\gamma_{\min}/(2B_0)\}$. Since $B_0$ is at most the number of edges of a path times $f(1)$, and $\gamma_{\min}$ is inverse-polynomial, $1/\eps$ is polynomially bounded. If $H$ has a multicoloured clique, then $\OPT(B_0)=L_0$ and the algorithm must return a path of length~$L_0$. Otherwise every path other than the fallback has charge at least $B_0+\gamma_{\min}>(1+\eps)B_0$ by Remark~\ref{rem:gap}, and the algorithm returns the fallback, of length $3L_0$. This decides \textsc{Multicolored Clique} in fixed-parameter time.
\end{proof}

The same fallback construction gives a standard gap statement for the length objective under the exact budget.

\begin{corollary}\label{cor:inapprox}
Let $f$ be a fixed tariff with $L_f>0$ and $0<\delta<1$. Unless $\FPT=\W1$, there is no algorithm with running time $g(\wr)\poly(N)$ that, on feasible unit-length instances, returns a path of charge at most $B$ and length at most $N^{1-\delta}\cdot\OPT(B)$.
\end{corollary}
\begin{proof}
First decide any source instance with an empty colour class $V_i$ or an empty edge set $E_{ij}$ as a no-instance, without constructing a fallback. For the remaining sources, let $S$ be the size of the instance of Theorem~\ref{thm:hard}, with common path length $L_0$, and add a fallback of $\mu L_0$ edges bearing one new label, for an odd $\mu=\Theta(S^b)$ with $b>2/\delta$. Here $B_0\ge f(1)>0$, so the fallback is feasible, and the width is at most $\max\{\wr,1\}$. Since $L_0\le S$, the number of added edges is $O(S^{b+1})$; allowing for binary vertex and label identifiers gives the coarse bound $N=O(S^{b+2})$. For all sufficiently large $S$,
\[
 \frac{\mu}{N^{1-\delta}}=\Omega\!\left(S^{\delta b+2\delta-2}\right)>1.
\]
If the source graph has a multicoloured clique, then $\OPT(B_0)=L_0$ and the approximation guarantee forces the returned path to have length less than $\mu L_0$; it must therefore lie in the original branch. Conversely, any exactly budget-feasible path in that branch encodes a clique by Theorem~\ref{thm:hard}. If no clique exists, the fallback is the only feasible branch. Inspecting the returned branch thus decides \textsc{Multicolored Clique} in fixed-parameter time. The finitely many smaller values of $S$ can be handled by exhaustive search.
\end{proof}

\section{Few labels}\label{sec:few}

The total number $q$ of labels and the number $\qrep$ of labels occurring on more than one edge are upper bounds on live width. They are easier to interpret but much more restrictive, and the reduction of Section~\ref{sec:hard} does not bound them. For constraints on the number of occurrences of each colour, $\W1$-hardness in the number of colours is known by a many-one reduction~\cite{BentertKN26}; that reduction prescribes a separate bound for every colour, whereas here a single budget bounds a sum of concave charges.

\begin{proposition}\label{prop:qxp}
For per-label tariffs and rational lengths, \textsc{Shortest Budgeted Path} can be solved in time $(T+1)^{q}\poly(N)$. In the common-tariff model it can be solved in time $(T+1)^{\qrep+1}\poly(N)$.
\end{proposition}
\begin{proof}
Store the last label and the run counts of all labels; the charge is a function of this vector, so it suffices to keep the minimum length for each vector. In the common-tariff model store the counts of the repeated labels and the number $n_1$ of edges traversed whose label occurs on no other edge; the charge is determined by these and equals the charge of the repeated labels plus $n_1f(1)$.
\end{proof}

For the lower bound we use a weaker reducibility. A \emph{disjunctive reduction} maps an instance $(x,k)$ in polynomial time to a list of instances $(y_1,p_1),\dots,(y_n,p_n)$ with $p_i\le\beta(k)$ for a computable $\beta$, such that $(x,k)$ is a yes-instance if and only if some $(y_i,p_i)$ is. A fixed-parameter algorithm for the target yields one for the source, and if $\beta(k)=O(k)$, an algorithm with running time $g(p)N^{o(p)}$ for the target contradicts ETH when the source is \textsc{Multicolored Clique}.

\begin{theorem}\label{thm:few}
Let $f$ be a fixed tariff on a polynomial grid with $L_f>0$. There is a disjunctive reduction from \textsc{Multicolored Clique} with $k$ colours to \textsc{Concave Run Budget} for $f$ on instances with at most $2k+1$ labels. The number of queries, their individual bit sizes, and their construction time are polynomial in the source input size, with exponents depending only on $f$ and independent of $k$. Hence feasibility is $\W1$-hard in $q$ and in $\qrep$ under disjunctive reductions, and has no $g(q)N^{o(q)}$ algorithm unless ETH fails.
\end{theorem}
\begin{proof}
Use the source notation of Section~\ref{sec:hard}, with $k\ge2$. If any colour class $V_i$ or edge set $E_{ij}$ is empty, output the single query consisting of one $s$--$t$ edge and budget zero; this is a no-instance because $L_f>0$ implies $f(1)>0$. Otherwise let $r_1<\dots<r_z$ be a $k$-strict code from Lemma~\ref{lem:codes}, and put $U=r_z+1$. In particular, $U\le C(kz)^e$ for constants $C,e$ depending only on $f$. The labels are $c_i,d_i$ for each colour $i$ and one separator $\zeta$. For a colour $i$ and an index $x$ let
\[
   W_i(x)=(\zeta c_i)^{r_x}(\zeta d_i)^{U-r_x}\zeta .
\]
Build the chain of gadgets of Theorem~\ref{thm:hard} with the words $W_i(x)$ in $\mathcal V_i$ and $W_i(x)W_j(y)$ in $\mathcal E_{ij}$, and label all selector and bridge edges with $\zeta$. On a path that selects the indices $x_{i,1},\dots,x_{i,k}$ for colour~$i$, the label $c_i$ has $s_i=\sum_mr_{x_{i,m}}$ runs, $d_i$ has $kU-s_i$ runs, and $\zeta$ has exactly $k^2U+1$ runs, since the $k^2U$ symbols other than $\zeta$ are isolated and the path begins and ends with~$\zeta$. With $\psi(t)=f(t)+f(kU-t)$ and $Z=f(k^2U+1)$,
\[
   \Phi(P)=Z+\sum_{i=1}^k\psi(s_i)\ \ge\ Z+V(P),\qquad V(P):=\frac1k\sum_{i=1}^k\sum_{m=1}^k\psi\bigl(kr_{x_{i,m}}\bigr).
\]
Indeed $\psi(s_i)-\frac1k\sum_m\psi(kr_{x_{i,m}})$ is the sum of $E_k(x_{i,1},\dots,x_{i,k})$ and of the Jensen gap of $f$ at the points $k(U-r_{x_{i,m}})$; both are nonnegative by Lemma~\ref{lem:gap}, and the first is positive unless all indices agree. So $\Phi(P)=Z+V(P)$ exactly for the paths that encode a multicoloured clique.

The value $V(P)$ depends on the clique, which is why a single budget does not suffice. Give the first edge of every word $W_i(x)$ the length $c_0-\psi(kr_x)/k$, where $c_0=\max_x\psi(kr_x)/k$, and all other edges length zero, so that $\ell(P)=k^2c_0-V(P)$. Let $M_0=kD_f(kU)$. All lengths and all values $V(P)$ are multiples of $1/M_0$, and $0\le V(P)\le2kf(kU)$. For each of the polynomially many multiples $b$ of $1/M_0$ in this range consider the two-budget question
\[
  \Phi(P)\le Z+b\quad\text{and}\quad \ell(P)\le k^2c_0-b .
\]
A path satisfying both has $\sum_i\psi(s_i)\le b\le V(P)\le\sum_i\psi(s_i)$ and therefore encodes a clique; conversely a clique path satisfies the question for $b=V(P)$.

It remains to express each two-budget question as an instance of feasibility without new labels. Let $\Lambda_b=M_0(k^2c_0-b)$; if $\Lambda_b<0$ the question is discarded. Form the DAG with vertices $(v,\lambda)$, $0\le\lambda\le\Lambda_b$ an integer, an edge from $(u,\lambda)$ to $(v,\lambda+M_0\ell(e))$ with the label of $e$ for every edge $e=(u,v)$ whose two endpoints lie in this range, source $(s,0)$, and a new target joined by a $\zeta$-labelled edge from every $(t,\lambda)$, and trim it. Its paths correspond to the paths of the original graph with $\ell(P)\le k^2c_0-b$, with the same label word followed by one $\zeta$, which prolongs the final run of $\zeta$. The budget is $Z+b$.

For the size bound, write $D=D_f(kU)$. There are at most $1+\lfloor2kM_0f(kU)\rfloor$ queries, and $c_0\le2f(kU)/k$ gives $\Lambda_b\le2kM_0f(kU)$. Since $M_0=kD$ and concavity gives $f(kU)\le kUf(1)$, both bounds are at most $1+2k^3DUf(1)$. The polynomial-grid hypothesis bounds $D$ by a fixed polynomial in $kU$, and $U\le C(kz)^e$. Before layering, the selection gadgets have $O(k^2z^2U)$ vertices and edges; the layered construction multiplies this by at most $1+\Lambda_b$ and adds a target. Rational arithmetic and label identifiers have polynomial bit length, including the value $Z=f(k^2U+1)$. Thus the number of queries, each query's bit size, and the total construction time are $(kz)^{O_f(1)}$, with an exponent independent of $k$.
\end{proof}

A many-one reduction would have to combine the queries, whose budgets $Z+b$ differ, over a bounded alphabet; we do not know how to do this. For membership, the plan representation gives a short certificate.

\begin{proposition}\label{prop:wp}
For per-label tariffs, \textsc{Concave Run Budget} parameterized by $q$, and also by $\qrep$, belongs to $\W{\mathrm P}$.
\end{proposition}
\begin{proof}
We use the characterisation of $\W{\mathrm P}$ by nondeterministic algorithms that run in fixed-parameter time and make $O(g(p)\log N)$ nondeterministic binary choices~\cite{ChenFG05}. Guess for every label $c$ either ``off'' or one of its at most $T$ supports $(\beta_c,\lambda_c)$; this takes $O(q\log(T+1))$ bits. Delete the edges with labels that are off, and compute a path that minimises the sum of $\lambda_c$ over its run starts, by a shortest path computation on the product with the last label. Accept if this minimum plus $\sum_c\beta_c$ over all labels that are not off is at most~$B$. If the algorithm accepts, the path found has $\Phi(P)\le\sum_{c\in\lab(P)}(\beta_c+\lambda_cr_c(P))\le B$, since supports dominate the tariffs and fees are nonnegative. Conversely, for a feasible path guess ``off'' for unused labels and a support attaining $f_c(r_c(P))$ for the used ones. For $\qrep$, guess only for repeated labels; a label on a single edge is charged $f_c(1)$ on that edge.
\end{proof}

Whether the problem is in $\W{\mathrm P}$, or in $\W1$, when parameterized by $\wr$ is open; the $\XP$ algorithms do not provide short certificates.

\section{Binary lengths: a numerical source of hardness}\label{sec:binary}

All hardness results so far concern feasibility, where numbers play no role beyond the tariff itself. \textsc{Shortest Budgeted Path} with rational lengths in binary has a second, purely numerical source of hardness, which appears already at width one for tariffs without a polynomial grid.

\begin{proposition}\label{prop:harmonic}
For the harmonic tariff $H$, deciding whether a path with $\Phi(P)\le B$ and $\ell(P)\le\Lambda$ exists is $\NP$-complete on series-parallel DAGs of maximum degree three with live width one and nonnegative rational lengths in binary.
\end{proposition}
\begin{proof}
We reduce from \textsc{Exact Cover by 3-Sets}~\cite{GareyJ79}: given a set $X$ of $3n$ elements and triples $S_1,\dots,S_m\subseteq X$, is $X$ the disjoint union of $n$ of the triples? Assign to each $e\in X$ a distinct prime $p_e>\max\{m,3\}$ of polynomial magnitude. A \emph{harmonic block} with $r$ runs is a new private label in $r$ runs separated by $r-1$ fresh labels; its charge is $H_r+r-1$. For each triple $S_i$ build two parallel branches: the \emph{exclude} branch is the concatenation of harmonic blocks with $p_e-1$ runs for $e\in S_i$, the \emph{include} branch that of blocks with $p_e$ runs. Join the gadgets of consecutive triples by fresh bridges. Including $S_i$ adds
\[
   \nu_i=3+\sum_{e\in S_i}\frac1{p_e}<4
\]
to the charge. Give one edge of each exclude branch the length $4$, one edge of each include branch the length $4-\nu_i$, and all other edges length zero. If $C$ is the charge of the all-exclude path and $I$ the set of included triples, then $\Phi(P)=C+\sum_{i\in I}\nu_i$ and $\ell(P)=4m-\sum_{i\in I}\nu_i$. With $\theta=3n+\sum_{e\in X}1/p_e$, $B=C+\theta$ and $\Lambda=4m-\theta$, both bounds hold if and only if $\sum_{i\in I}\nu_i=\theta$. If $m_e$ denotes the number of included triples containing~$e$, this reads
\[
   3(|I|-n)+\sum_{e\in X}\frac{m_e-1}{p_e}=0 .
\]
Multiplying by $\prod_ep_e$ and reducing modulo $p_e$ gives $m_e\equiv1\pmod{p_e}$, hence $m_e=1$ since $0\le m_e\le m<p_e$. So $I$ is an exact cover, and conversely an exact cover satisfies the equation. Only the private label of the current block is live at any vertex. Block sizes and bit lengths are polynomial.
\end{proof}

This is the knapsack mechanism of restricted shortest paths, carried by the arithmetic of harmonic numbers; the lengths have a common denominator of exponential size. It disappears with unary lengths, where Theorem~\ref{thm:plans}(2) with the baseline keys is polynomial at bounded width, and it does not occur for tariffs on a polynomial grid, for which rational lengths are in $\XP$ (Corollary~\ref{cor:examples}). Table~\ref{tab:landscape} summarises the picture for a fixed common tariff.

\begin{table}[t]
\centering\small
\begin{tabular}{@{}p{43mm}p{50mm}p{50mm}@{}}
\toprule
Fixed common tariff & Feasibility & Shortest path, rational lengths in binary\\
\midrule
Bounded $J_f$ (fixed monetary grid, caps, finite menus) & FPT in $\wr$ (Prop.~\ref{prop:grid}) & FPT in $\wr$ (Prop.~\ref{prop:grid})\\[2pt]
Effectively subpolynomial $J_f$ & FPT in $\wr$ (Cor.~\ref{cor:fpt}) & FPT in $\wr$ on a polynomial grid (Cor.~\ref{cor:fpt})\\[2pt]
Rounded twins $F_M$ & FPT in $\wr$ (Prop.~\ref{prop:twins}) & FPT in $\wr$ (Prop.~\ref{prop:twins})\\[2pt]
$L_f>0$ & $\W1$-hard in $\wr$, in $\XP$ (Thm.~\ref{thm:hard}) & $\W1$-hard in $\wr$\\[2pt]
$L_f>0$ on a polynomial grid, e.g.\ $F$, $G_\alpha$ & $\W1$-hard in $\wr$; hard in $q$ under disjunctive reductions (Thms.~\ref{thm:residue}, \ref{thm:few}) & $\W1$-hard in $\wr$, in $\XP$ (Cor.~\ref{cor:examples})\\[2pt]
Harmonic $H$ & $\W1$-hard in $\wr$, in $\XP$ & $\NP$-hard at $\wl=1$ (Prop.~\ref{prop:harmonic})\\
\bottomrule
\end{tabular}
\caption{Complexity for a fixed tariff shared by all labels. Feasibility results extend to unit and polynomially bounded integer lengths.}
\label{tab:landscape}
\end{table}

\section{Open problems}\label{sec:open}

\paragraph{The oscillating regime.} Theorem~\ref{thm:main} leaves open the tariffs with $L_f=0<U_f$. For such a tariff $T^*_f$ is not polynomially bounded (Lemma~\ref{lem:growth}), so strict codes of polynomial size exist only for infinitely many, sparsely placed lengths~$z$, and padding a source instance to the next good length may cost superpolynomial size. On the other hand the plan algorithm is slow at the horizons where $J_f(T)\ge T^{\delta}$. We do not know whether such tariffs can be fixed-parameter tractable, or whether hardness can be shown by a reduction that does not rely on a single uniform code.

\paragraph{Effectiveness.} The remaining $U_f=0$ cases are those without an effective subpolynomial bound. Can the effectiveness requirement be weakened, or can a different algorithm yield a computable parameter dependence for these tariffs? Together with the oscillating regime, this prevents Theorem~\ref{thm:main} from being a complete dichotomy.

\paragraph{Classes.} Is \textsc{Concave Run Budget}, for a hard tariff, complete for $\W1$ when parameterized by $\wr$, or does it lie higher? We do not even know membership in $\W{\mathrm P}$.

\paragraph{Few labels.} Is feasibility $\W1$-hard in $q$ under many-one reductions, and is the grid hypothesis of Theorem~\ref{thm:few} necessary? The harmonic tariff is a natural test case.

\paragraph{Width constants.} Theorem~\ref{thm:hard} gives width $2k+1$ and Theorem~\ref{thm:residue} gives $k+1$ on polynomial grids. Can $k+O(1)$ be achieved without a grid? This would require paying $C-f(kr_x)/k$, or a substitute, with tariff values only.

\paragraph{Accuracy.} Corollary~\ref{cor:accuracy} separates $g(\wr,1/\eps)\poly(N)$ from $g(\wr)\poly(N,1/\eps)$ for polynomially separated tariffs. The best joint dependence on $\wr$ and $1/\eps$, and the case of exponentially small separation such as geometric marginals, remain open.

\section*{Code availability}
Source code is available at \url{https://github.com/ndehouche/concave-run-tariffs}.


\begin{thebibliography}{99}\small

\bibitem{AsahiroJMO22} Y.~Asahiro, J.~Jansson, E.~Miyano, H.~Ono. Upper and lower degree-constrained graph orientation with minimum penalty. \emph{Theoretical Computer Science} 900:53--78, 2022.

\bibitem{BarrettJM00} C.~Barrett, R.~Jacob, M.~Marathe. Formal-language-constrained path problems. \emph{SIAM Journal on Computing} 30(3):809--837, 2000. \href{https://doi.org/10.1137/S0097539798337716}{doi:\nolinkurl{10.1137/S0097539798337716}}.

\bibitem{BentertKN22} M.~Bentert, L.~Kellerhals, R.~Niedermeier. The structural complexity landscape of finding balance-fair shortest paths. \emph{Theoretical Computer Science} 933:149--162, 2022. \href{https://doi.org/10.1016/j.tcs.2022.08.023}{doi:\nolinkurl{10.1016/j.tcs.2022.08.023}}.

\bibitem{BentertKN26} M.~Bentert, L.~Kellerhals, R.~Niedermeier. Fair short paths in vertex-colored graphs. \emph{Theoretical Computer Science} 1085:116201, 2026. \href{https://doi.org/10.1016/j.tcs.2026.116201}{doi:\nolinkurl{10.1016/j.tcs.2026.116201}}. Conference version: \emph{AAAI 2023}, 37(10):12346--12354, 2023. \href{https://doi.org/10.1609/aaai.v37i10.26455}{doi:\nolinkurl{10.1609/aaai.v37i10.26455}}.

\bibitem{BlancoEtAl16} M.~Blanco, R.~Bornd\"orfer, N.~D.~Ho\`ang, A.~Kaier, T.~Schlechte, S.~Schlobach. The shortest path problem with crossing costs. ZIB-Report 16-70, Zuse Institute Berlin, 2016. \href{https://opus4.kobv.de/opus4-zib/files/6124/ZR-16-70.pdf}{Report PDF}; \href{https://nbn-resolving.org/urn:nbn:de:0297-zib-61240}{urn:nbn:de:0297-zib-61240}.

\bibitem{BlancoEtAl17} M.~Blanco, R.~Bornd\"orfer, N.~D.~Ho\`ang, A.~Kaier, P.~Maristany de las Casas, T.~Schlechte, S.~Schlobach. Cost projection methods for the shortest path problem with crossing costs. In \emph{ATMOS 2017}, OASIcs 59, 15:1--15:14, 2017. \href{https://doi.org/10.4230/OASIcs.ATMOS.2017.15}{doi:\nolinkurl{10.4230/OASIcs.ATMOS.2017.15}}.

\bibitem{Bodlaender93} H.~L.~Bodlaender. A tourist guide through treewidth. \emph{Acta Cybernetica} 11(1--2):1--21, 1993.

\bibitem{BohmovaEtAl18} K.~B\"ohmov\'a, L.~H\"afliger, M.~Mihal\'ak, T.~Pr\"oger, G.~Sacomoto, M.-F.~Sagot. Computing and listing $st$-paths in public transportation networks. \emph{Theory of Computing Systems} 62(3):600--621, 2018. \href{https://doi.org/10.1007/s00224-016-9747-4}{doi:\nolinkurl{10.1007/s00224-016-9747-4}}.

\bibitem{BredereckEtAl20} R.~Bredereck, P.~Faliszewski, R.~Niedermeier, P.~Skowron, N.~Talmon. Mixed integer programming with convex/concave constraints: fixed-parameter tractability and applications to multicovering and voting. \emph{Theoretical Computer Science} 814:86--105, 2020. \href{https://doi.org/10.1016/j.tcs.2020.01.017}{doi:\nolinkurl{10.1016/j.tcs.2020.01.017}}.

\bibitem{BroersmaLWZ05} H.~Broersma, X.~Li, G.~J.~Woeginger, S.~Zhang. Paths and cycles in colored graphs. \emph{Australasian Journal of Combinatorics} 31:299--311, 2005.

\bibitem{BumpusM23} B.~M.~Bumpus, K.~Meeks. Edge exploration of temporal graphs. \emph{Algorithmica} 85(3):688--716, 2023. \href{https://doi.org/10.1007/s00453-022-01018-7}{doi:\nolinkurl{10.1007/s00453-022-01018-7}}.

\bibitem{CardinalD08} J.~Cardinal, C.~Dumeunier. Set covering problems with general objective functions. Preprint, \href{https://arxiv.org/abs/0802.2184}{arXiv:0802.2184}, 2008.

\bibitem{ChenHKX06} J.~Chen, X.~Huang, I.~A.~Kanj, G.~Xia. Strong computational lower bounds via parameterized complexity. \emph{Journal of Computer and System Sciences} 72(8):1346--1367, 2006. \href{https://doi.org/10.1016/j.jcss.2006.04.007}{doi:\nolinkurl{10.1016/j.jcss.2006.04.007}}.

\bibitem{ChenFG05} Y.~Chen, J.~Flum, M.~Grohe. Machine-based methods in parameterized complexity theory. \emph{Theoretical Computer Science} 339(2--3):167--199, 2005. \href{https://doi.org/10.1016/j.tcs.2005.02.003}{doi:\nolinkurl{10.1016/j.tcs.2005.02.003}}.

\bibitem{CyganEtAl15} M.~Cygan, F.~V.~Fomin, {\L}.~Kowalik, D.~Lokshtanov, D.~Marx, M.~Pilipczuk, M.~Pilipczuk, S.~Saurabh. \emph{Parameterized Algorithms}. Springer, 2015. \href{https://doi.org/10.1007/978-3-319-21275-3}{doi:\nolinkurl{10.1007/978-3-319-21275-3}}.

\bibitem{DavotGK24} T.~Davot, R.~Giroudeau, J.-C.~K\"onig. On the shared transportation problem: computational hardness and exact approach. \emph{International Journal of Foundations of Computer Science} 35(6):741--756, 2024. \href{https://doi.org/10.1142/S0129054123410058}{doi:\nolinkurl{10.1142/S0129054123410058}}.

\bibitem{Dehouche20} N.~Dehouche. The $k$-interchange-constrained diameter of a transit network: a connectedness indicator that accounts for travel convenience. \emph{Transportation Letters} 12(3):197--201, 2020. \href{https://doi.org/10.1080/19427867.2018.1564987}{doi:\nolinkurl{10.1080/19427867.2018.1564987}}.

\bibitem{DellingPW15} D.~Delling, T.~Pajor, R.~F.~Werneck. Round-based public transit routing. \emph{Transportation Science} 49(3):591--604, 2015. \href{https://doi.org/10.1287/trsc.2014.0534}{doi:\nolinkurl{10.1287/trsc.2014.0534}}.

\bibitem{DowneyF13} R.~G.~Downey, M.~R.~Fellows. \emph{Fundamentals of Parameterized Complexity}. Springer, 2013. \href{https://doi.org/10.1007/978-1-4471-5559-1}{doi:\nolinkurl{10.1007/978-1-4471-5559-1}}.

\bibitem{EnrightHLM26} J.~Enright, S.~D.~Hand, L.~Larios-Jones, K.~Meeks. Families of tractable problems with respect to vertex-interval-membership width and its generalisations. In \emph{SAND 2026}, LIPIcs 373, 4:1--4:17, 2026. \href{https://doi.org/10.4230/LIPIcs.SAND.2026.4}{doi:\nolinkurl{10.4230/LIPIcs.SAND.2026.4}}.

\bibitem{EulerB19} R.~Euler, R.~Bornd\"orfer. A graph- and monoid-based framework for price-sensitive routing in local public transportation networks. In \emph{ATMOS 2019}, OASIcs 75, 12:1--12:15, 2019. \href{https://doi.org/10.4230/OASIcs.ATMOS.2019.12}{doi:\nolinkurl{10.4230/OASIcs.ATMOS.2019.12}}.

\bibitem{EulerLB24} R.~Euler, N.~Lindner, R.~Bornd\"orfer. Price optimal routing in public transportation. \emph{EURO Journal on Transportation and Logistics} 13:100128, 2024. \href{https://doi.org/10.1016/j.ejtl.2024.100128}{doi:\nolinkurl{10.1016/j.ejtl.2024.100128}}.

\bibitem{FellowsGK10} M.~R.~Fellows, J.~Guo, I.~Kanj. The parameterized complexity of some minimum label problems. \emph{Journal of Computer and System Sciences} 76(8):727--740, 2010. \href{https://doi.org/10.1016/j.jcss.2010.02.012}{doi:\nolinkurl{10.1016/j.jcss.2010.02.012}}.

\bibitem{FellowsHRV09} M.~R.~Fellows, D.~Hermelin, F.~Rosamond, S.~Vialette. On the parameterized complexity of multiple-interval graph problems. \emph{Theoretical Computer Science} 410(1):53--61, 2009. \href{https://doi.org/10.1016/j.tcs.2008.09.065}{doi:\nolinkurl{10.1016/j.tcs.2008.09.065}}.

\bibitem{GabrielB97} S.~A.~Gabriel, D.~Bernstein. The traffic equilibrium problem with nonadditive path costs. \emph{Transportation Science} 31(4):337--348, 1997. \href{https://doi.org/10.1287/trsc.31.4.337}{doi:\nolinkurl{10.1287/trsc.31.4.337}}.

\bibitem{GamvrosGR12} I.~Gamvros, L.~Gouveia, S.~Raghavan. Reload cost trees and network design. \emph{Networks} 59(4):365--379, 2012. \href{https://doi.org/10.1002/net.20443}{doi:\nolinkurl{10.1002/net.20443}}.

\bibitem{GareyJ79} M.~R.~Garey, D.~S.~Johnson. \emph{Computers and Intractability: A Guide to the Theory of NP-Completeness}. W.~H.~Freeman, 1979.

\bibitem{GoelTW10} G.~Goel, P.~Tripathi, L.~Wang. Combinatorial problems with discounted price functions in multi-agent systems. In \emph{FSTTCS 2010}, LIPIcs 8, pp.~436--446, 2010. \href{https://doi.org/10.4230/LIPIcs.FSTTCS.2010.436}{doi:\nolinkurl{10.4230/LIPIcs.FSTTCS.2010.436}}.

\bibitem{GourvesLMM10} L.~Gourv\`es, A.~Lyra, C.~A.~Martinhon, J.~Monnot. The minimum reload $s$--$t$ path, trail and walk problems. \emph{Discrete Applied Mathematics} 158(13):1404--1417, 2010. \href{https://doi.org/10.1016/j.dam.2010.03.009}{doi:\nolinkurl{10.1016/j.dam.2010.03.009}}.

\bibitem{GozupekOPSS17} D.~G\"oz\"upek, S.~\"Ozkan, C.~Paul, I.~Sau, M.~Shalom. Parameterized complexity of the MinCCA problem on graphs of bounded decomposability. \emph{Theoretical Computer Science} 690:91--103, 2017. \href{https://doi.org/10.1016/j.tcs.2017.06.013}{doi:\nolinkurl{10.1016/j.tcs.2017.06.013}}.

\bibitem{GozupekSSZ16} D.~G\"oz\"upek, H.~Shachnai, M.~Shalom, S.~Zaks. Constructing minimum changeover cost arborescenses in bounded treewidth graphs. \emph{Theoretical Computer Science} 621:22--36, 2016. \href{https://doi.org/10.1016/j.tcs.2016.01.022}{doi:\nolinkurl{10.1016/j.tcs.2016.01.022}}.

\bibitem{Hassin92} R.~Hassin. Approximation schemes for the restricted shortest path problem. \emph{Mathematics of Operations Research} 17(1):36--42, 1992. \href{https://doi.org/10.1287/moor.17.1.36}{doi:\nolinkurl{10.1287/moor.17.1.36}}.

\bibitem{HassinMS07} R.~Hassin, J.~Monnot, D.~Segev. Approximation algorithms and hardness results for labeled connectivity problems. \emph{Journal of Combinatorial Optimization} 14(4):437--453, 2007. \href{https://doi.org/10.1007/s10878-007-9044-x}{doi:\nolinkurl{10.1007/s10878-007-9044-x}}.

\bibitem{IrnichD05} S.~Irnich, G.~Desaulniers. Shortest path problems with resource constraints. In \emph{Column Generation}, pp.~33--65. Springer, 2005.

\bibitem{Kinnersley92} N.~G.~Kinnersley. The vertex separation number of a graph equals its path-width. \emph{Information Processing Letters} 42(6):345--350, 1992. \href{https://doi.org/10.1016/0020-0190(92)90234-M}{doi:\nolinkurl{10.1016/0020-0190(92)90234-M}}.

\bibitem{LorenzR01} D.~H.~Lorenz, D.~Raz. A simple efficient approximation scheme for the restricted shortest path problem. \emph{Operations Research Letters} 28(5):213--219, 2001. \href{https://doi.org/10.1016/S0167-6377(01)00069-4}{doi:\nolinkurl{10.1016/S0167-6377(01)00069-4}}.

\bibitem{MagnantiS04} T.~L.~Magnanti, D.~Stratila. Separable concave optimization approximately equals piecewise linear optimization. In \emph{IPCO 2004}, LNCS 3064, pp.~234--243. Springer, 2004. \href{https://doi.org/10.1007/978-3-540-25960-2_18}{doi:\nolinkurl{10.1007/978-3-540-25960-2_18}}. Full version: \href{https://arxiv.org/abs/1201.3148}{arXiv:1201.3148}, 2012.

\bibitem{MagnantiS12} T.~L.~Magnanti, D.~Stratila. Strongly polynomial primal-dual algorithms for concave cost combinatorial optimization problems. Preprint, \href{https://arxiv.org/abs/1202.2654}{arXiv:1202.2654}, 2012.

\bibitem{Parmentier19} A.~Parmentier. Algorithms for non-linear and stochastic resource constrained shortest path. \emph{Mathematical Methods of Operations Research} 89(2):281--317, 2019. \href{https://doi.org/10.1007/s00186-018-0649-x}{doi:\nolinkurl{10.1007/s00186-018-0649-x}}.

\bibitem{SchobelU22} A.~Sch\"obel, R.~Urban. The cheapest ticket problem in public transport. \emph{Transportation Science} 56(6):1432--1451, 2022. \href{https://doi.org/10.1287/trsc.2022.1138}{doi:\nolinkurl{10.1287/trsc.2022.1138}}.

\bibitem{TsaggourisZ09} G.~Tsaggouris, C.~Zaroliagis. Multiobjective optimization: improved FPTAS for shortest paths and non-linear objectives with applications. \emph{Theory of Computing Systems} 45(1):162--186, 2009. \href{https://doi.org/10.1007/s00224-007-9096-4}{doi:\nolinkurl{10.1007/s00224-007-9096-4}}.

\bibitem{WirthS01} H.-C.~Wirth, J.~Steffan. Reload cost problems: minimum diameter spanning tree. \emph{Discrete Applied Mathematics} 113(1):73--85, 2001. \href{https://doi.org/10.1016/S0166-218X(00)00392-9}{doi:\nolinkurl{10.1016/S0166-218X(00)00392-9}}.

\bibitem{ZhangSS16} Y.~Zhang, Z.-J.~M.~Shen, S.~Song. Parametric search for the bi-attribute concave shortest path problem. \emph{Transportation Research Part B: Methodological} 94:150--168, 2016. \href{https://doi.org/10.1016/j.trb.2016.09.009}{doi:\nolinkurl{10.1016/j.trb.2016.09.009}}.

\end{thebibliography}
\end{document}